%% file: conference_101719.tex
\documentclass[10pt, conference, letterpaper]{IEEEtran}
\IEEEoverridecommandlockouts
\usepackage{cite}
\usepackage{amsmath,amssymb,amsfonts}
\usepackage{algorithmic}
\usepackage{graphicx}
\usepackage{textcomp}
\usepackage{xcolor}
\usepackage{subcaption} 
\usepackage{caption}  
\usepackage{float} 
\usepackage{booktabs}  
\usepackage[table]{xcolor} 
\usepackage{siunitx}  
\usepackage{makecell}  
\usepackage{multirow}  
\usepackage{hyperref} 
\usepackage{xurl} 
\usepackage{placeins} 

\def\BibTeX{{\rm B\kern-.05em{\sc i\kern-.025em b}\kern-.08em
    T\kern-.1667em\lower.7ex\hbox{E}\kern-.125emX}}
\begin{document}

\title{FedJam: Multimodal Federated Learning Framework for Jamming Detection%
\thanks{\textsuperscript{1}All artifacts are available at \url{https://github.com/panitsasi/fedJam}.}
}

\author{
    \IEEEauthorblockN{Ioannis Panitsas\textsuperscript{*}, Iason Ofeidis\textsuperscript{*}, Leandros Tassiulas}
    \IEEEauthorblockA{
        Department of Electrical and Computer Engineering, Yale University
    }
    \thanks{\textsuperscript{*}These authors contributed equally to this work.}
}

\maketitle

\begin{abstract}

Jamming attacks pose a critical threat to wireless networks, yet existing detection methods remain largely unimodal, centralized and resource-intensive, limiting their performance, scalability, and deployment feasibility, respectively. To address these limitations, we present \textit{FedJam}, a multimodal Federated Learning (FL) framework for on-device jamming detection and classification. \textit{FedJam} locally fuses spectrograms and cross-layer network Key Performance Indicators (KPIs) using a lightweight dual-encoder architecture with an integrated fusion module and multimodal projection head, that enables privacy-preserving training and inference without transmitting raw data. We prototype and deploy \textit{FedJam} on a wireless experimental testbed and evaluate it using the first, over-the-air multimodal dataset comprising synchronized samples across benign and three distinct jamming attack types. \textit{FedJam} outperforms state-of-the-art unimodal baselines by up to \textit{15\%} in accuracy, while requiring \textit{60\% fewer} communication rounds to converge, and maintains low resource utilization. Its advantage is especially pronounced in realistic scenarios, where it remains extremely robust under heterogeneous data distributions across devices. 

\end{abstract}


\section{Introduction}

Wireless networks serve as the backbone of modern communication systems, enabling a wide range of services. However, their open and shared medium exposes them to physical-layer security threats; particularly jamming attacks, where adversaries intentionally interfere with legitimate wireless signals to disrupt communication~\cite{jamming_survey}. Once largely associated with military operations, jamming has emerged as a broader threat vector, with recent incidents disrupting commercial aviation via GPS interference~\cite{gps_jamming_baltic} and disabling home security systems through Wi-Fi jamming during burglaries~\cite{wifi_jamming_glendale}. The accessibility of low-cost Software-Defined Radios (SDRs) and open-source software has drastically lowered the barrier to launching such attacks. For instance, USB dongles costing under \$20 can be configured to emit disruptive signals, enabling denial-of-service attacks with minimal technical expertise~\cite{cheap_sdr_jamming}.

In response, several techniques have been introduced to detect and categorize such attack vectors. Conventional approaches rely on energy-based methods, monitoring signal strength anomalies or disruptions in Key Performance Indicators (KPIs)~\cite{jamming_survey}, while state-of-the-art approaches increasingly rely on Machine Learning (ML) to analyze In-phase and Quadrature (I/Q) samples and detect interference patterns. For instance, common approaches transform I/Q streams into spectrogram representations and employ computer vision models to detect jamming activity ~\cite{spectrogram_3, spectrogram_4}.

Despite their demonstrated effectiveness, existing ML-based methods continue to face challenges in terms of computational efficiency, scalability, and real-world deployability. First, many proposed detection models are computationally intensive, employing deep and complex architectures that prioritize accuracy over efficiency, thereby constraining their deployability in resource-constrained environments~\cite{spectrogram_5, spectrogram_6}. 
Second, most prior systems are designed as centralized solutions~\cite{spectrogram_1, spectrogram_2, kpi1} that rely on collecting large volumes of raw data (e.g., I/Q streams) from sensors or base stations and transmitting them to a central server for analysis and decision-making. This process not only imposes substantial burdens on compute and network resources, but also raises scalability concerns. In addition, transmitting such data to centralized servers introduces additional security risks, as raw I/Q samples and KPIs may reveal sensitive information, including user identities, locations, and behavioral patterns~\cite{iq_security}. Centralization further amplifies the risk of interception, tampering, and adversarial attacks, such as model poisoning~\cite{adversarial_attacks}.

\begin{figure}[t]
  \centering
  \includegraphics[width=\columnwidth]{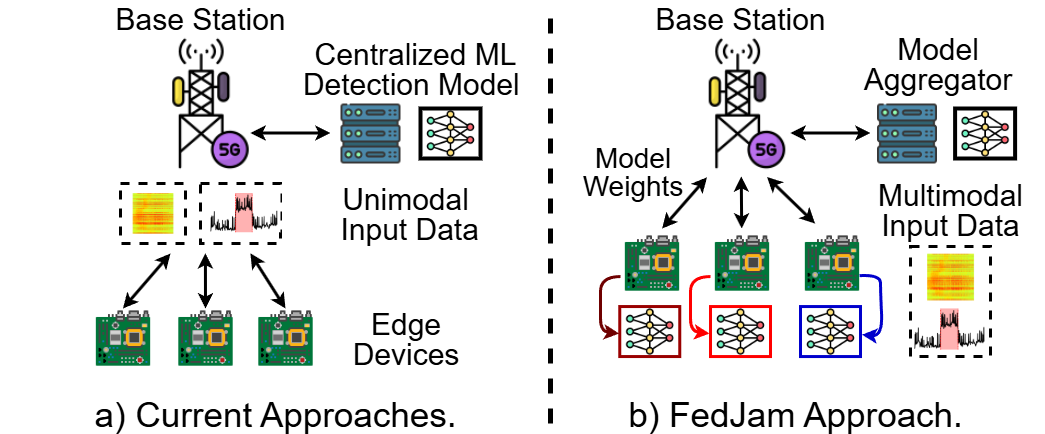} 
  \caption{\textit{FedJam} enables collaborative, on-device learning over multimodal data without exposing raw private measurements.}
  \label{fig:intro}
\end{figure}

Beyond these challenges, prior detection solutions focus solely on unimodal sources, overlooking complementary information naturally present within the network across diverse modalities, which can limit their ability to form a holistic view of the network state~\cite{spectrogram_1, kpi1, kpi2}. This lack of cross-modal integration can result to misdetections, especially when sophisticated attackers mask their activity in one modality while leaving subtle traces in another~\cite{sumalatha2024comprehensive}. Although multimodal architectures have demonstrated state-of-the-art performance in multiple domains such as vision-language understanding~\cite{multimodal_sota}, their application to wireless security remains largely unexplored. A major obstacle is the lack of publicly available datasets that capture time-aligned multimodal network information, which are essential for developing and evaluating effective multimodal detection systems. Existing datasets are predominantly synthetic and fail to reflect real-world complexity, limiting the robustness, practical relevance, and deployment feasibility of current approaches~\cite{spectrogram_5, spectrogram_6}.

To address these challenges, we introduce \textit{FedJam}, a privacy-preserving, multimodal Federated Learning (FL) framework for on-device jamming detection and classification that operates seamlessly on resource-constrained devices across diverse wireless backhauls. As illustrated in Fig.~\ref{fig:intro}, \textit{FedJam} shifts from centralized detection pipelines to a collaborative learning paradigm. It operates across distributed, sensing-capable devices, performing on-device training and inference without sharing raw data. \textit{FedJam} leverages a lightweight architecture to jointly process multimodal inputs (spectrograms and multi-channel time series KPIs), enabling the extraction of both frequency-domain characteristics and performance degradation patterns. It consists of six key components: (i) a time alignment module that synchronizes spectrograms and KPI sequences over fixed time windows, (ii) an adaptive temporal downsampler that reduces the resolution of KPI sequences to enable trade-offs between efficiency and accuracy under resource constraints, (iii) a spectrum encoder that extracts spectro-temporal features from spectrograms, (iv) a KPI encoder that captures temporal dynamics from cross-layer KPIs, (v) a fusion module that integrates modality-specific embeddings into a unified network state representation, and (vi) a lightweight projection head that generates task-specific outputs. To evaluate its real-world effectiveness, we prototype and deploy \textit{FedJam} on a wireless experimental testbed, demonstrating its practicality and robustness. Extensive evaluations show that \textit{FedJam} outperforms state-of-the-art unimodal baselines by up to 15\% in accuracy while incurring minimal resource overhead. The main contributions of this work are:

\begin{itemize}

\item We propose \textit{FedJam}, a multimodal FL framework for jamming detection and classification that operates efficiently in resource-constrained environments with low computational and communication overhead.

\item We design a lightweight multimodal architecture that combines spectral and temporal network features into a unified representation for accurate interference detection.

\item We collect a large-scale, over-the-air multimodal dataset from a real-world wireless testbed under diverse and controlled interference scenarios.

\item We prototype and deploy \textit{FedJam} on a real experimental testbed, demonstrating its superiority over state-of-the-art baselines, with its advantage being especially pronounced in realistic non-IID (Independent and Identically Distributed) scenarios.

\item We release all artifacts (including dataset, code, and deployment scripts)\textsuperscript{1} to foster reproducibility and enable further research.

\end{itemize}

The rest of the paper is structured as follows. Section~\ref{sec:related} reviews related work. Section~\ref{sec:fedjam} introduces our proposed framework. Section~\ref{sec:evaluation} outlines the experimental setup and presents the results. Finally, Section~\ref{sec:conclusion} concludes the paper.


\section{Related Work}
\label{sec:related}

\noindent \textbf{Centralized Detection Approaches.} Centralized ML-based detection systems have dominated early research on jamming detection and classification~\cite{spectrogram_1, spectrogram_2, spectrogram_3, spectrogram_4, spectrogram_5, spectrogram_6, iq_samples_1_deep_learning, iq_samples_2_deep_learning, iq_samples_3_deep_learning}. These approaches typically rely on offloading continuous, high-volume data, such as raw I/Q samples or KPI streams, to a centralized server for training and inference~\cite{iq_samples_1_deep_learning, iq_samples_2_deep_learning, kpi1, kpi2}. To improve detection accuracy, many of these systems employ deep and complex models, such as computer vision-based architectures~\cite{spectrogram_1, spectrogram_2, spectrogram_3, spectrogram_4, spectrogram_5, spectrogram_6}. Despite their high accuracy, these centralized solutions remain challenging to deploy in practice due to their significant compute and communication demands, as well as data privacy concerns.

\noindent  \textbf{Distributed and Collaborative Detection Approaches.} To improve scalability and preserve data privacy, recent efforts have explored FL-based techniques~\cite{federated_learning_attacks, federated_learning_spectrogram_1,federated_learning_spectrogram_2,federated_learning_spectrogram_3,federated_learning_spectrogram_4,federated_learning_spectrogram_5}. These approaches enable multiple devices to collaboratively train and fine-tune a shared detection model, while also supporting local inference without exposing raw private data to an untrusted centralized entity. However, they still suffer from key limitations. Existing approaches rely on heavyweight ML models for jamming detection, assuming abundant computational resources—often unrealistic in IoT and mobile scenarios. Moreover, they depend on single-source inputs, overlooking complementary data such as time-series metrics that offer critical context for accurate detection.

\noindent \textbf{Unimodal Detection Approaches.} Ongoing research on ML-based jamming detection has predominantly relied on unimodal learning, using either spectrum-level representations~\cite{spectrogram_1,spectrogram_2,spectrogram_3,spectrogram_4,spectrogram_5,spectrogram_6, iq_samples_1_deep_learning, iq_samples_2_deep_learning, iq_samples_3_deep_learning} or network-level statistics~\cite{kpi1,kpi2,kpi3,kpi4}.
Spectrum-centric approaches transform raw I/Q samples into spectrograms and employ deep learning models \cite{spectrogram_1,spectrogram_2,spectrogram_3, spectrogram_4,spectrogram_5,spectrogram_6} to identify interference patterns. In contrast, models based on network- or device-level statistics, such as signal quality, block error rate, and packet loss ratio, typically rely on threshold-based logic or anomaly detection frameworks that learn the normal behavior of the network and flag deviations \cite{kpi1,kpi2, kpi3, kpi4}. However, relying on a unimodal perspective limits these approaches’ ability to capture a comprehensive view of the network, reducing their effectiveness against sophisticated attacks.

\noindent \textbf{Multimodal Federated Learning.} Multimodal FL integrates heterogeneous data sources, such as images, audio, and time-series signals, to capture complementary information and improve collaborative learning robustness. While this paradigm is gaining traction in fields such as healthcare, human activity recognition, and autonomous systems, its application to wireless security remains limited \cite{mmsurvey}. Despite its potential, existing multimodal FL approaches across domains still face two significant limitations. Firstly, they exhibit poor performance under non-IID conditions, an inherent challenge in real-world deployments where clients experience diverse environments and modality distributions~\cite{chen2022towards}. Secondly, they frequently overlook the increased computational cost of processing multiple data streams, often assuming that all clients have uniform and sufficient resources~\cite{Che2023Multimodal}. These limitations hinder their applicability in resource-constrained, heterogeneous edge scenarios, such as wireless jamming detection.

\section{FedJam}
\label{sec:fedjam}


In this section, we present \textit{FedJam}, a novel multimodal FL framework for on-device jamming detection and classification. Designed as an out-of-the-box solution, \textit{FedJam} operates seamlessly on resource-constrained devices and remains agnostic to both the wireless backhaul and the underlying learning and aggregation setup.

\subsection{System Overview}
\label{sec:fedjam_overview}

\begin{figure}
    \centering
    \includegraphics[width=\linewidth]{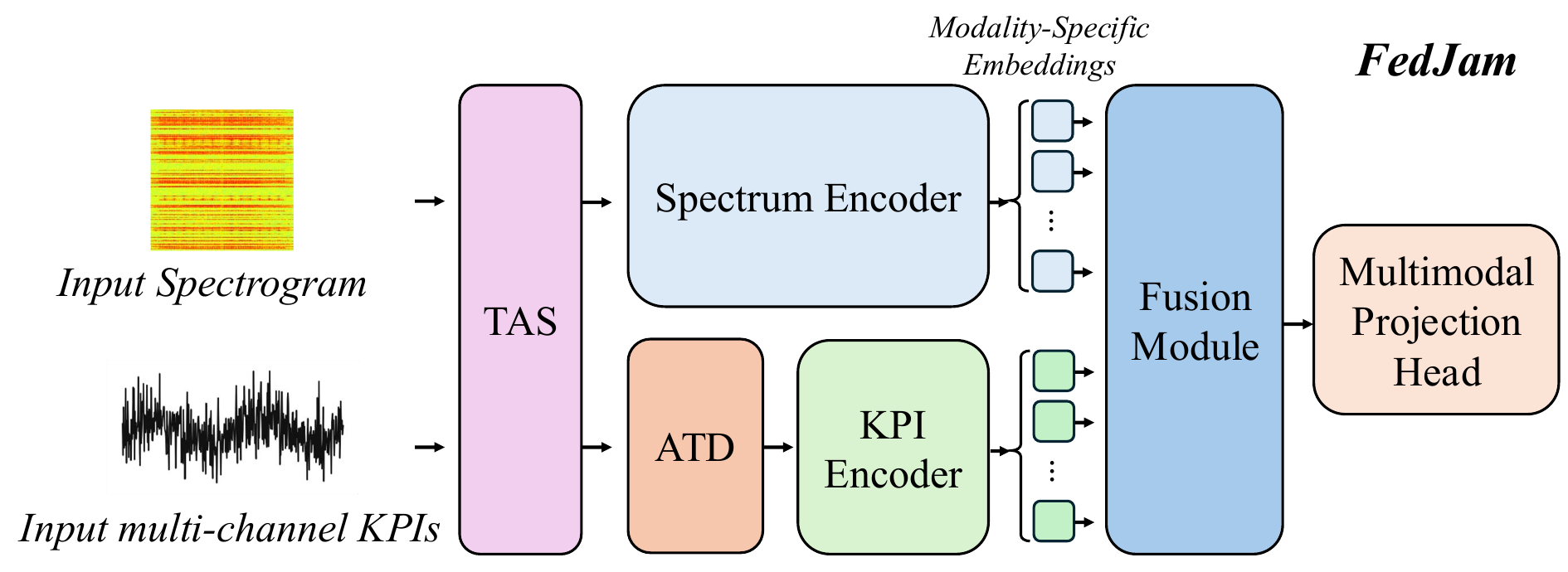}
    \caption{FedJam system architecture.}
    \label{fig:fedjam_architecture}
\end{figure}

\textit{FedJam} is a lightweight, multimodal framework for on-device jamming detection and classification. It processes two complementary input modalities collected over fixed time windows: spectrum activity, represented as multi-channel spectrogram images, and network-level KPIs, represented as multi-channel time-series sequences. As illustrated in Fig.~\ref{fig:fedjam_architecture}, \textit{FedJam} comprises six functional blocks: (i) a \textit{Time Alignment and Synchronization Module} that ensures both modalities are temporally aligned within each window; (ii) an \textit{Adaptive Temporal Downsampler} that reduces the resolution of KPI sequences based on device constraints, enabling efficient processing; (iii) a dual-branch multimodal encoder consisting of a \textit{Spectrum Encoder} and a \textit{KPI Encoder}, which operate in parallel to extract high-level features from each modality; (iv) a \textit{Fusion Module} that combines the resulting embeddings into a unified representation of the network state; and (v) a \textit{Multimodal Projection Head} that transforms this representation into task-specific outputs, such as multi-class labels indicating the presence and type of interference. \textit{FedJam} operates entirely at the edge: both training and inference are performed locally on each device. During training, each client updates its model using private data and periodically transmits only model parameters to a central server. The server aggregates these updates using a standard learning method, producing a new global model that is redistributed to the clients.

\textit{FedJam} is designed to operate seamlessly in real-world, heterogeneous environments. Its architecture is inherently agnostic to the underlying wireless backhaul, supporting deployment over diverse connectivity options such as Wi-Fi and 5G without requiring any modifications or tuning. Moreover, \textit{FedJam} is resilient to partial modality availability: if one input stream (e.g., spectrograms) is missing or degraded due to sensor faults, network constraints, or runtime disruptions, the system gracefully degrades to unimodal inference using only the available encoder and an adjusted projection head. This built-in fault tolerance ensures continued operation, robust decision-making, and reliable performance even under adverse network or device conditions.


\subsection{System Components}
\label{sec:fedjam_components}


Building on the high-level system overview, we now examine the internal structure of \textit{FedJam}, highlighting the main functionality and operational role of each core system module. More specifically, \textit{FedJam} consists of:

\noindent \textbf{(I) Time Alignment and Synchronization (TAS) Module.}
This module performs fine-grained temporal alignment of spectrogram and KPI modalities by partitioning both streams into fixed-length, synchronized observation windows. It ensures strict cross-modal temporal coherence, a prerequisite for effective multimodal feature extraction and downstream fusion. By enforcing consistency across modalities, TAS enables time-locked feature comparisons and facilitates accurate detection of transient interference events.

\noindent \textbf{(II) Adaptive Temporal Downsampler (ATD).} To reduce computation and memory usage on constrained devices, \textit{FedJam} incorporates an Adaptive Temporal Downsampler that reduces the granularity of KPI time-series inputs. It applies a lightweight downsampling, such as fixed-stride sampling or average pooling, based on device-specific constraints (e.g., latency or energy budget). This enables a tunable trade-off between accuracy and efficiency, allowing \textit{FedJam} to gracefully adapt to a wide range of resource profiles. The design is also extensible to learned or attention-guided sampling strategies.

\noindent \textbf{(III) Spectrum Encoder.}
To capture fine-grained physical-layer interference characteristics, the spectrum encoder processes spectrogram inputs that represent the local spectral activity observed over a given time window. It employs a compact visual backbone to extract salient frequency-domain and spatial features indicative of jamming behavior, such as anomalous energy bursts, narrowband spikes, or wideband noise patterns. By compressing high-dimensional spectrograms into compact embeddings, this module provides an efficient and expressive summary of spectral dynamics, enabling accurate  downstream processing.

\noindent \textbf{(IV) KPI Encoder.}  
Operating in parallel with the spectrum encoder, the KPI encoder processes multi-channel time-series KPIs that reflect the temporal evolution of network-layer behavior over the same observation window. It captures both intra-channel dynamics and inter-channel correlations that often reveal performance degradation caused by jamming, uncovering cross-layer patterns that remain invisible in the spectral domain alone. This complementary modality significantly strengthens the system's situational awareness, enabling the detection of subtle, delayed, or stealthy interference that may evade purely physical-layer analysis.

\noindent \textbf{(V) Fusion Module.}  
The fusion module integrates modality-specific embeddings from the spectrum and KPI encoders to construct a unified, context-aware representation of the network state. This cross-modal representation enables the system to leverage both fine-grained spectral patterns and broader temporal trends in network performance. Notably, the two modalities may exhibit asymmetric temporal importance: spectral features often capture short-lived, bursty interference, while KPI sequences reflect longer-term degradations or performance drift. By jointly encoding these perspectives, the fusion module enables the system to adaptively determine the relevance of each modality and effectively combine their contributions, producing a unified network state representation.

\noindent \textbf{(VI) Multimodal Projection Head Module.}  
The final module maps the fused representation to a task-specific output. While this module can be generalized to support a variety of downstream tasks, in our setting it produces a classification decision indicating the presence and type of interference. Designed to be lightweight and modular, it allows easy adaptation to alternative tasks with minimal architectural changes, enabling broad applicability across various scenarios.

\subsection{Training Workflow} 
\label{sec:training_workflow}

\textit{FedJam} adopts a synchronous FL setup to collaboratively train a multimodal model across multiple edge clients. Each client performs local updates on its private spectrogram and KPI data, and periodically transmits selected model parameters, namely those of the spectrum encoder, KPI encoder, and classifier, to a central server. To minimize communication costs, these updates may be compressed before transmission. The server aggregates the received updates using any aggregation method (e.g., FedAvg~\cite{fedavg}) and broadcasts the updated global model back to all clients. This process is repeated for a fixed number of communication rounds or until convergence, enabling the multimodal model to learn effectively from decentralized data without sharing raw samples.

\section{Experimental Setup and Data Collection}
\label{sec:exp_setup}

\subsection{Experimental Environment}

\begin{figure*}[ht]
    \centering
    \begin{subfigure}[b]{0.19\textwidth}
        \captionsetup{justification=centering, labelformat=empty}
        \includegraphics[height=3cm]{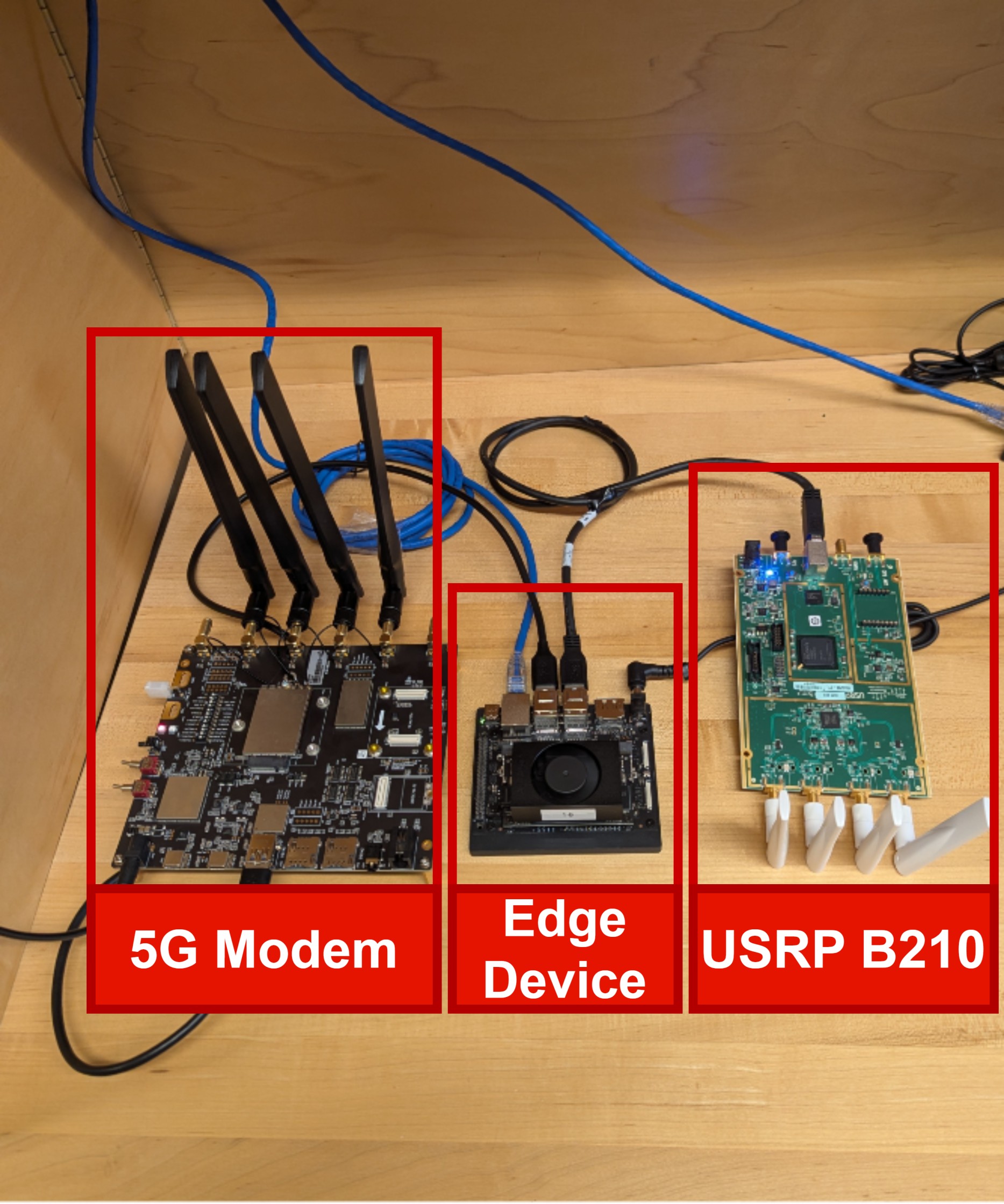}
        \caption{\hspace{-3.5em} (a) Edge device.}
        \label{fig:jetson_device}
    \end{subfigure}\hspace{-2em}
    \begin{subfigure}[b]{0.19\textwidth}
        \captionsetup{justification=centering, labelformat=empty}
        \includegraphics[height=3cm]{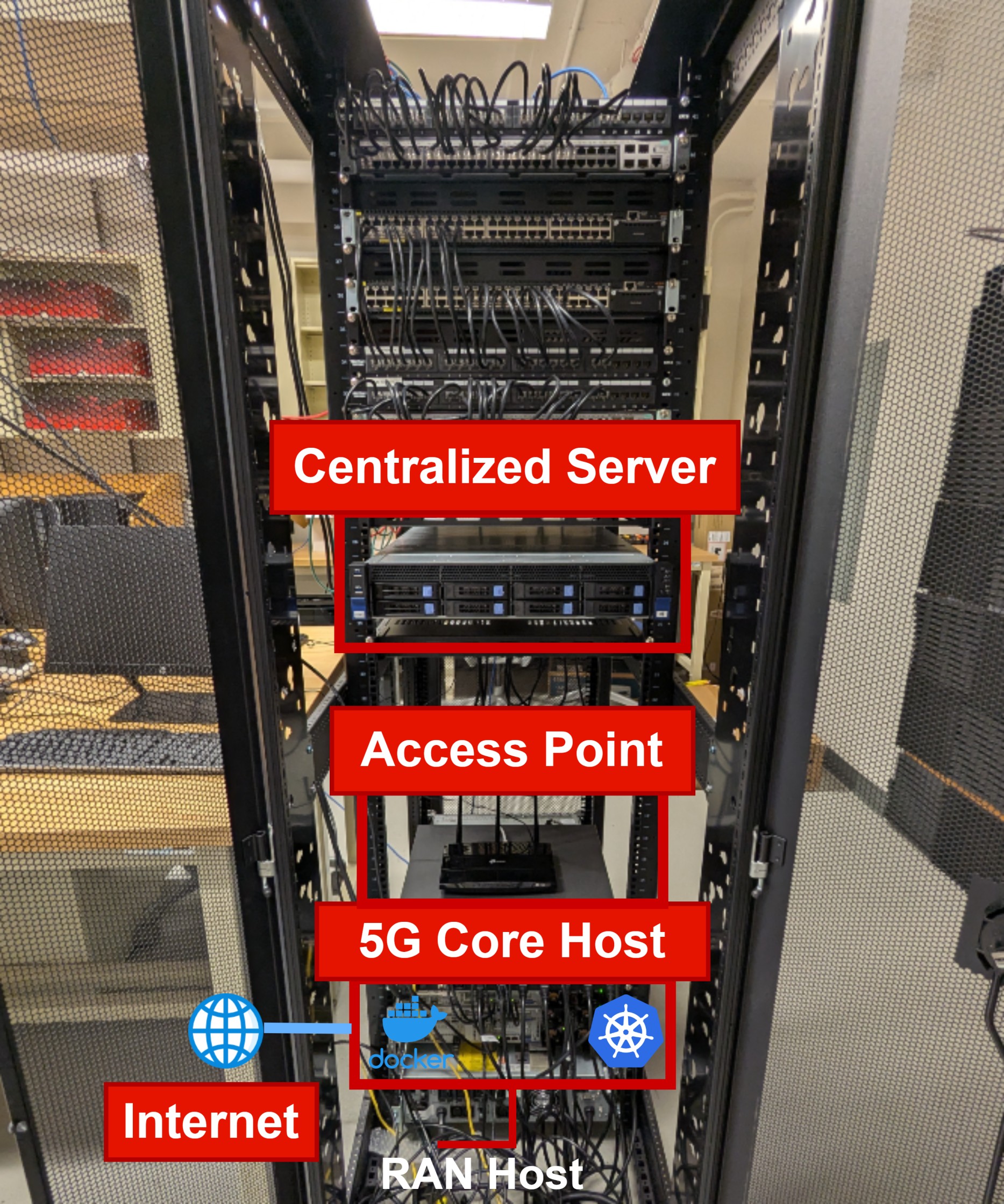}
        \caption{\hspace{-2.8em} (b) Centralized server.}
        \label{fig:rack_server}
    \end{subfigure}\hspace{-2.2em}
    \begin{subfigure}[b]{0.19\textwidth}
        \captionsetup{justification=centering, labelformat=empty}
        \includegraphics[height=3cm]{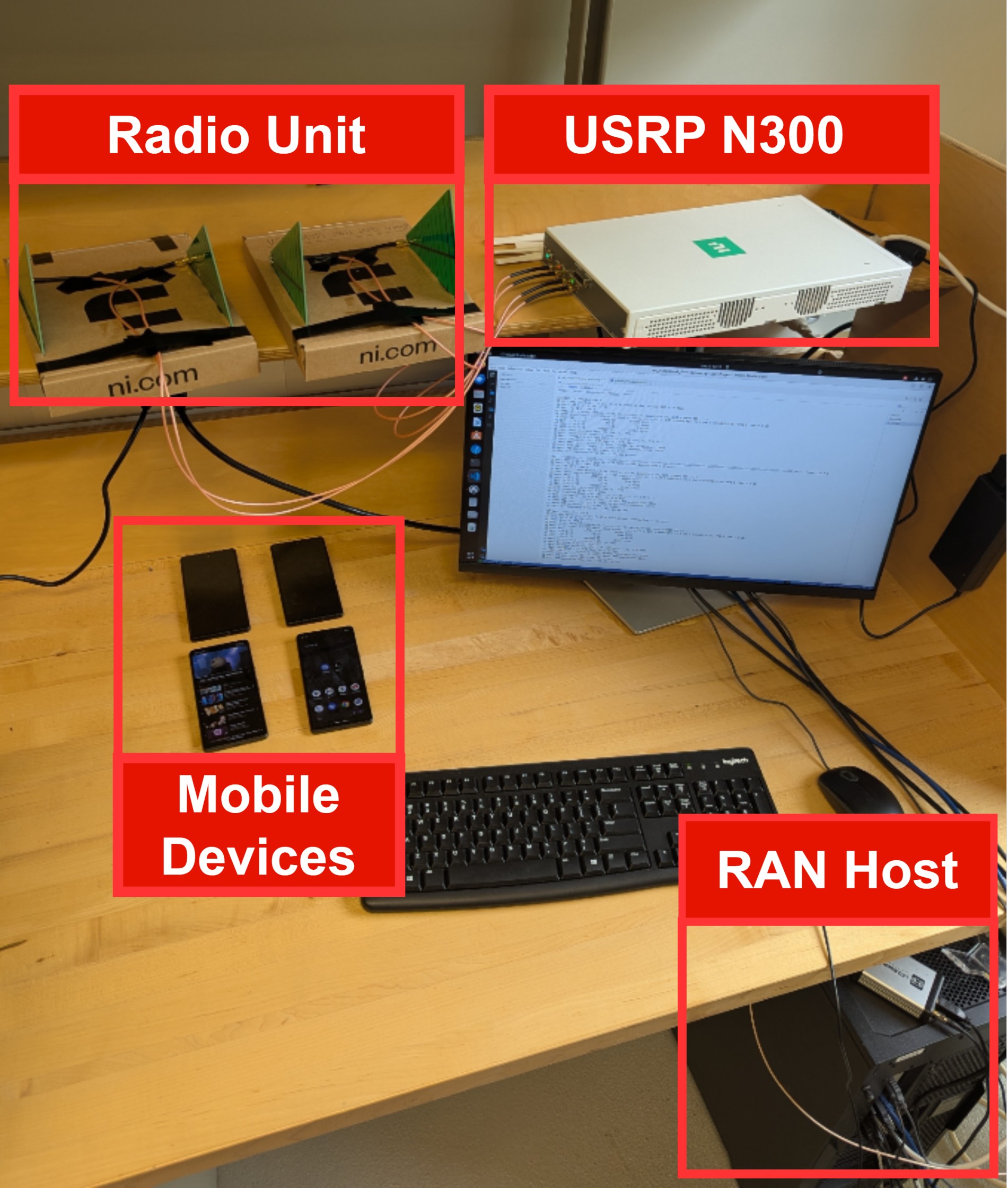}
        \caption{\hspace{-3em} (c) 5G testbed.}
        \label{fig:testbed}
    \end{subfigure}\hspace{-2em}
    \begin{subfigure}[b]{0.19\textwidth}
        \captionsetup{justification=centering, labelformat=empty}
        \includegraphics[height=3cm]{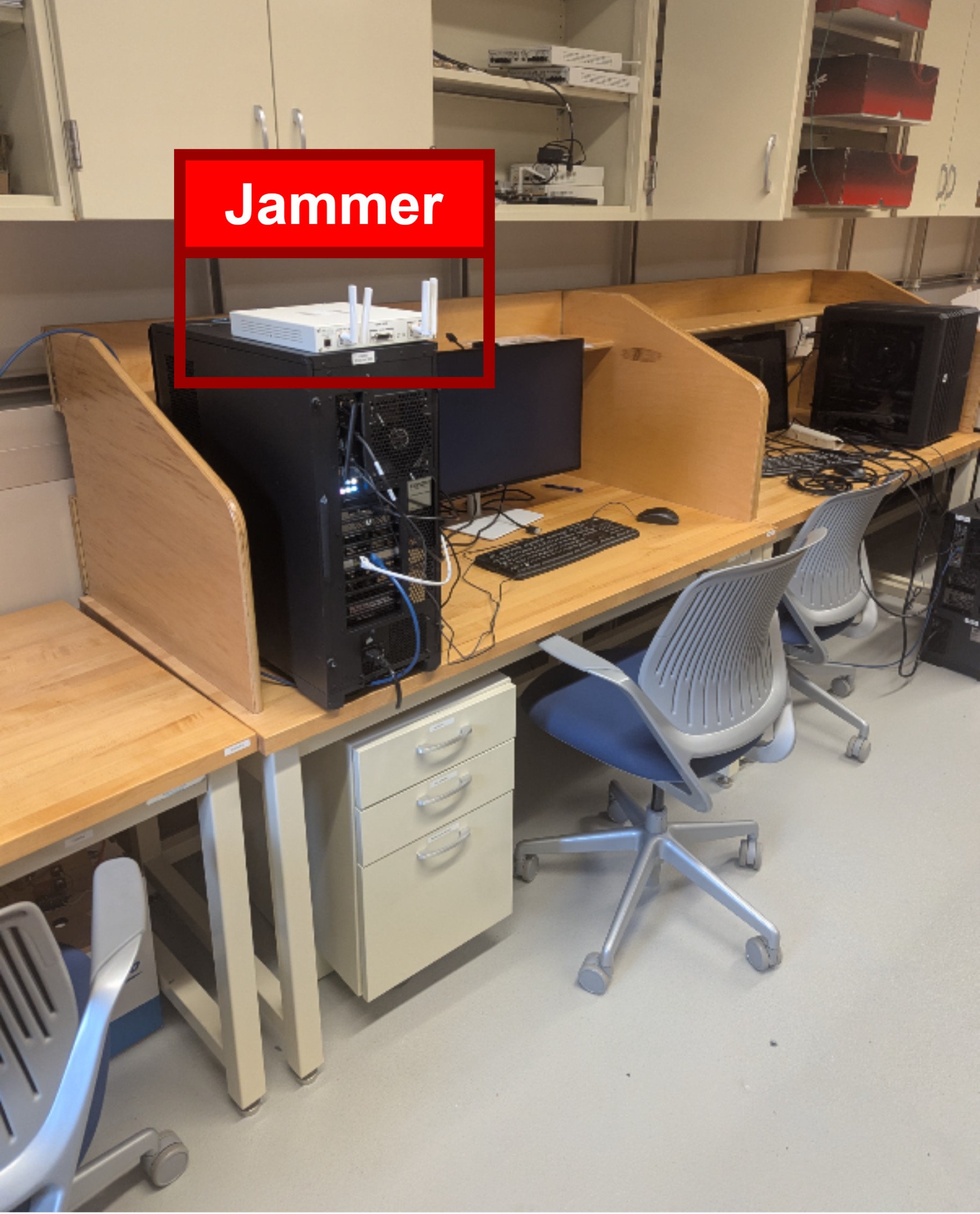}
        \caption{\hspace{-3.5em} (d) Jammer.}
        \label{fig:jammer}
    \end{subfigure}\hspace{-2.3em}
    \begin{subfigure}[b]{0.19\textwidth}
        \captionsetup{justification=centering, labelformat=empty}
        \includegraphics[height=3cm]{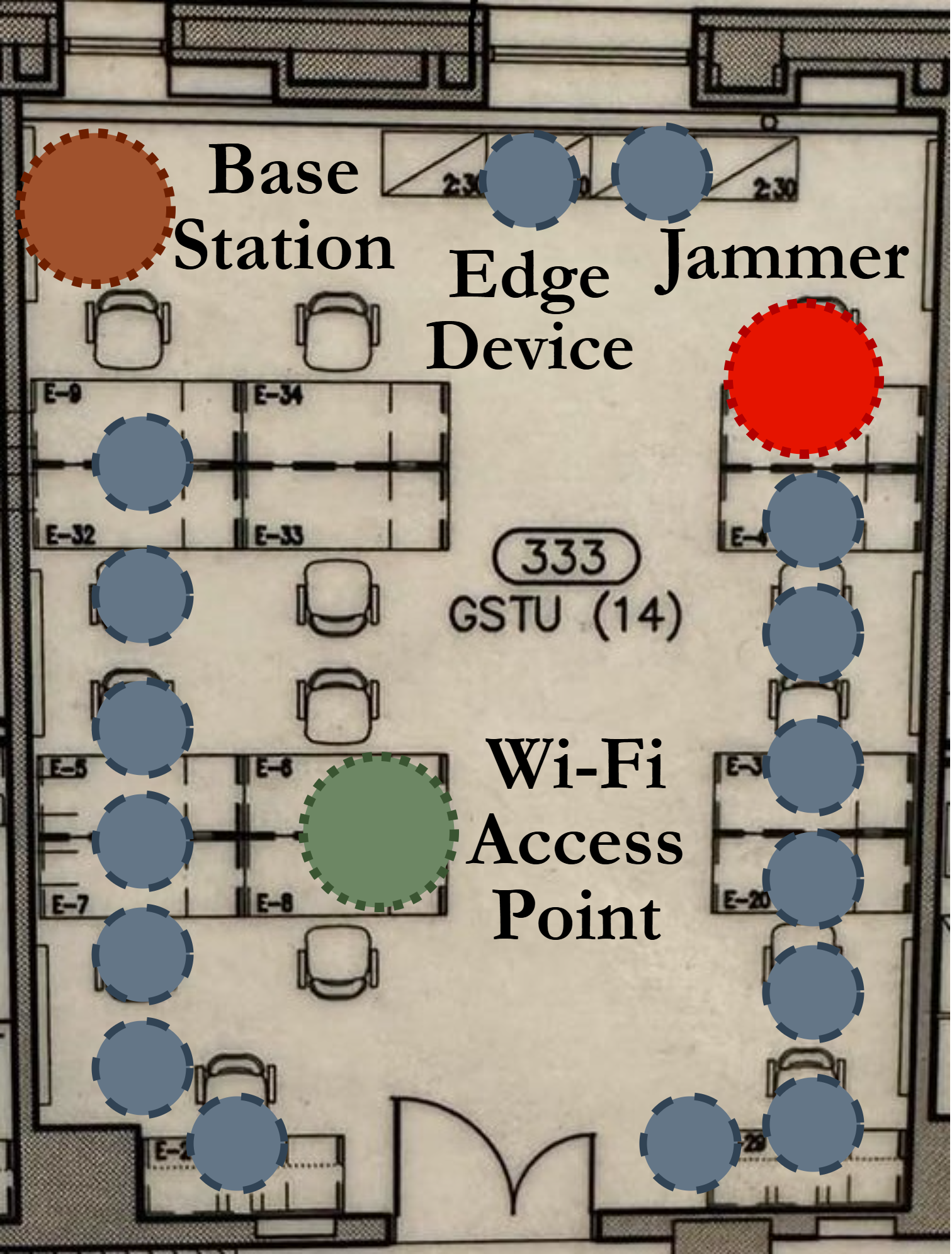}
        \caption{\hspace{-3.8em} (e) Lab layout.}
        \label{fig:lab_view}
    \end{subfigure}
    \caption{Experimental testbed components and lab layout.}
    \label{fig:testbed_overview}
\end{figure*}

\noindent \textbf{Experimental Testbed.} As illustrated in Fig.~\ref{fig:testbed_overview}, we deploy \textit{FedJam} on a wireless testbed consisting of 16 edge devices and a central server. Each edge device, shown in Fig.~\ref{fig:jetson_device}, is an NVIDIA Jetson Orin Nano running Ubuntu 22.04 and is equipped with a 6-core ARM Cortex CPU, a 1024-core NVIDIA Ampere GPU, 8 GB of shared system and GPU memory, and a 500 GB NVMe SSD. For  spectrum monitoring, the device interfaces with a USRP B210 SDR, via a dedicated USB 3.0 connection. In addition, each device connects to a Quectel RM520N-GL 5G modem, mounted on a 5G-M2-EVB-KIT evaluation board, via a dedicated USB 3.0 connection, with a programmable SIM card enabling 5G connectivity. Finally, the central server, shown in Fig.~\ref{fig:rack_server}, is equipped with a 32-core AMD Ryzen Threadripper PRO CPU, 504 GB of memory, 3.5 TB of storage, and four RTX A6000 GPUs. 

\noindent \textbf{Network Connectivity.} Connectivity between the edge devices and the central server is supported through both Wi-Fi and 5G. Wi-Fi connectivity is provided by a TP-Link Archer C7 router running OpenWRT firmware, operating exclusively in the 2.4 GHz band across the three non-overlapping 20 MHz channels (1, 6, and 11). 5G connectivity is enabled through a fully customized private 5G network operating in the n78 TDD band with 20 MHz bandwidth centered at 3.319 GHz, using a 5 ms radio frame structure with 7 downlink slots and 2 uplink slots at 30 kHz subcarrier spacing. This private 5G deployment is supported by a separate testbed developed in-house, shown in Fig.~\ref{fig:rack_server} and Fig.~\ref{fig:testbed}, featuring a  monolithic base station (gNB), connected to a USRP N300 SDR via a 10 Gbps fronthaul link, with the SDR further connected to 6 dBi directional antennas to enable over-the-air transmission. The gNB is further connected to the 5G core network via a dedicated 10 Gbps backhaul link, with all 3GPP-compliant Virtual Network Functions (VNFs) deployed as containerized services within a customized Kubernetes cluster. Both the Radio Access Network (RAN) and core network VNFs are implemented using the latest release of the OpenAirInterface (OAI) 5G software stack~\cite{oai}. 

\begin{figure}[t]
    \centering
    \begin{subfigure}[b]{0.24\columnwidth}
        \includegraphics[width=0.9\linewidth]{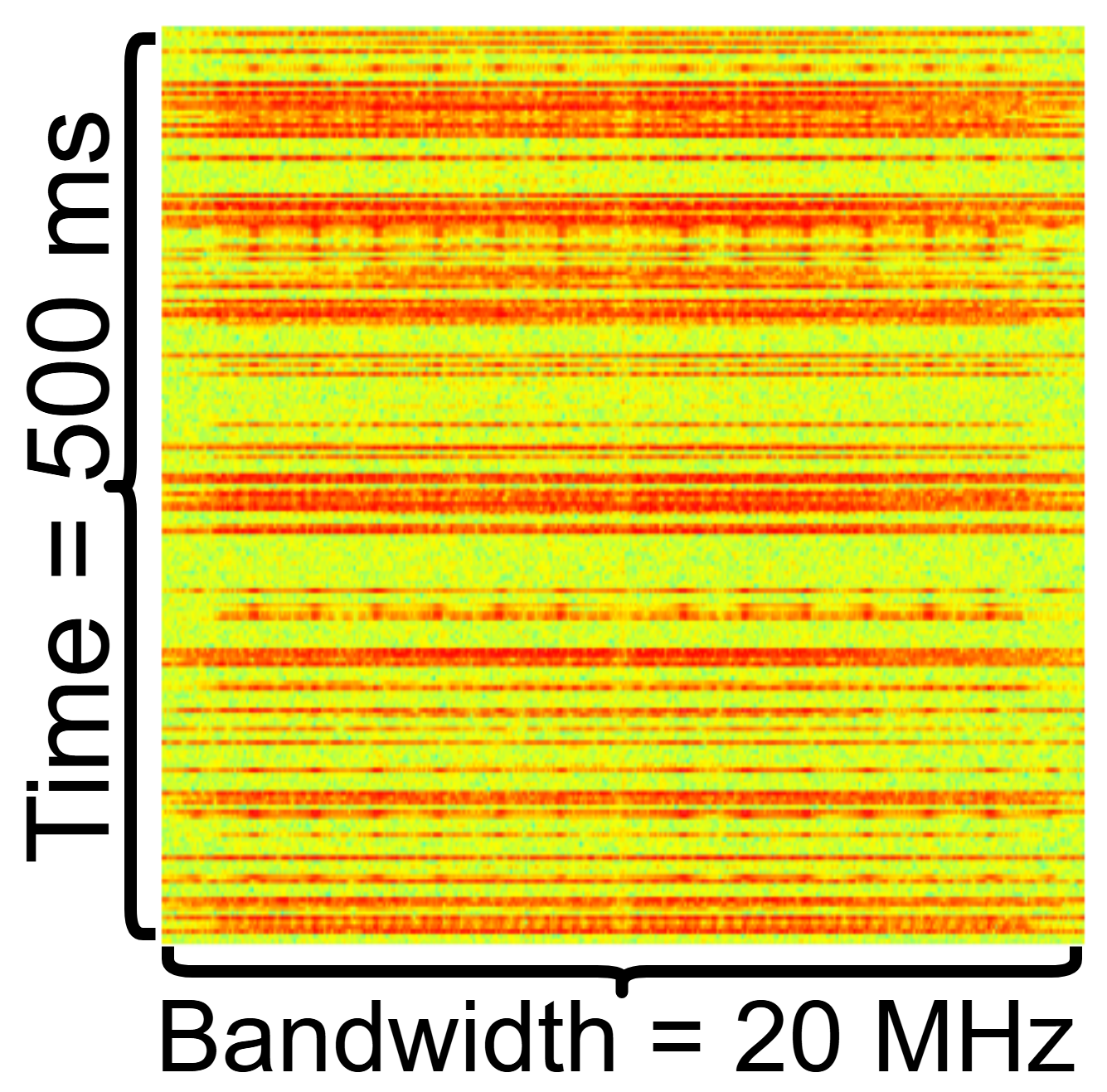}
        \caption{Benign.}
        \label{fig:spec_benign}
    \end{subfigure}
    \hfill
    \begin{subfigure}[b]{0.24\columnwidth}
        \includegraphics[width=0.9\linewidth]{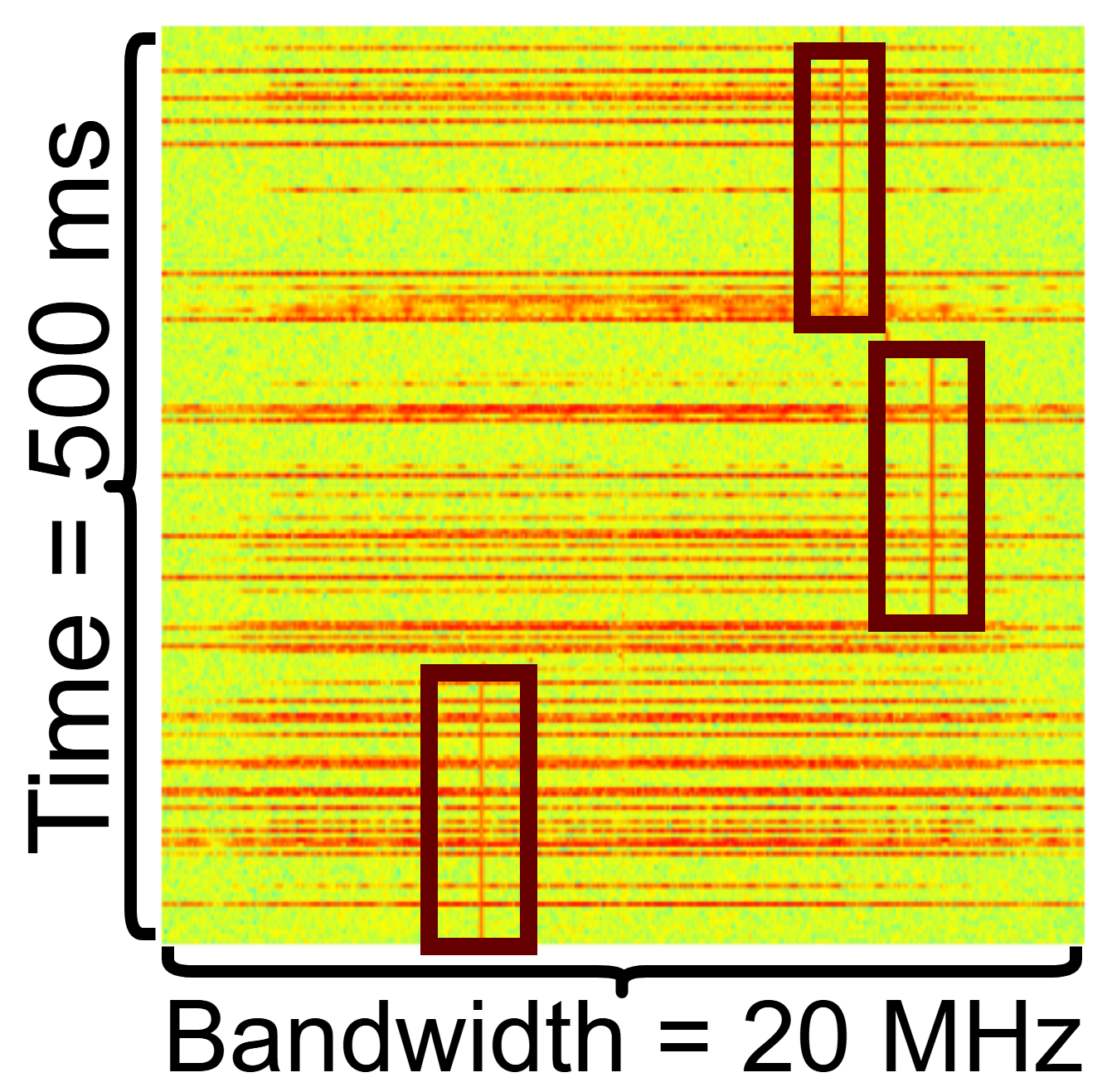}
        \caption{Single-tone.}
        \label{fig:spec_single_tone}
    \end{subfigure}
    \hfill
    \begin{subfigure}[b]{0.24\columnwidth}
        \includegraphics[width=0.9\linewidth]{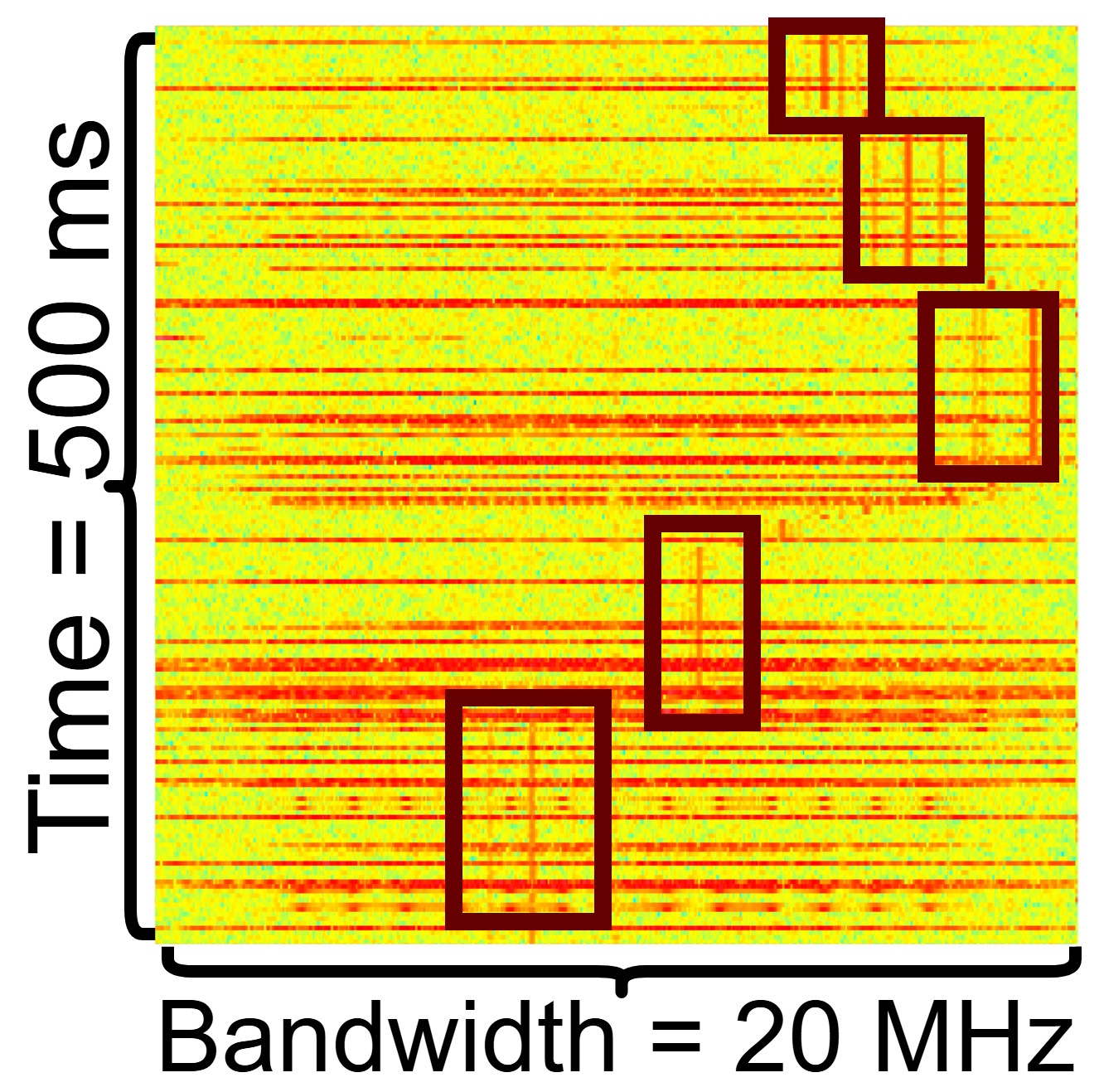}
        \caption{Pulse.}
        \label{fig:spec_pulse}
    \end{subfigure}
    \hfill
    \begin{subfigure}[b]{0.24\columnwidth}
        \includegraphics[width=0.9\linewidth]{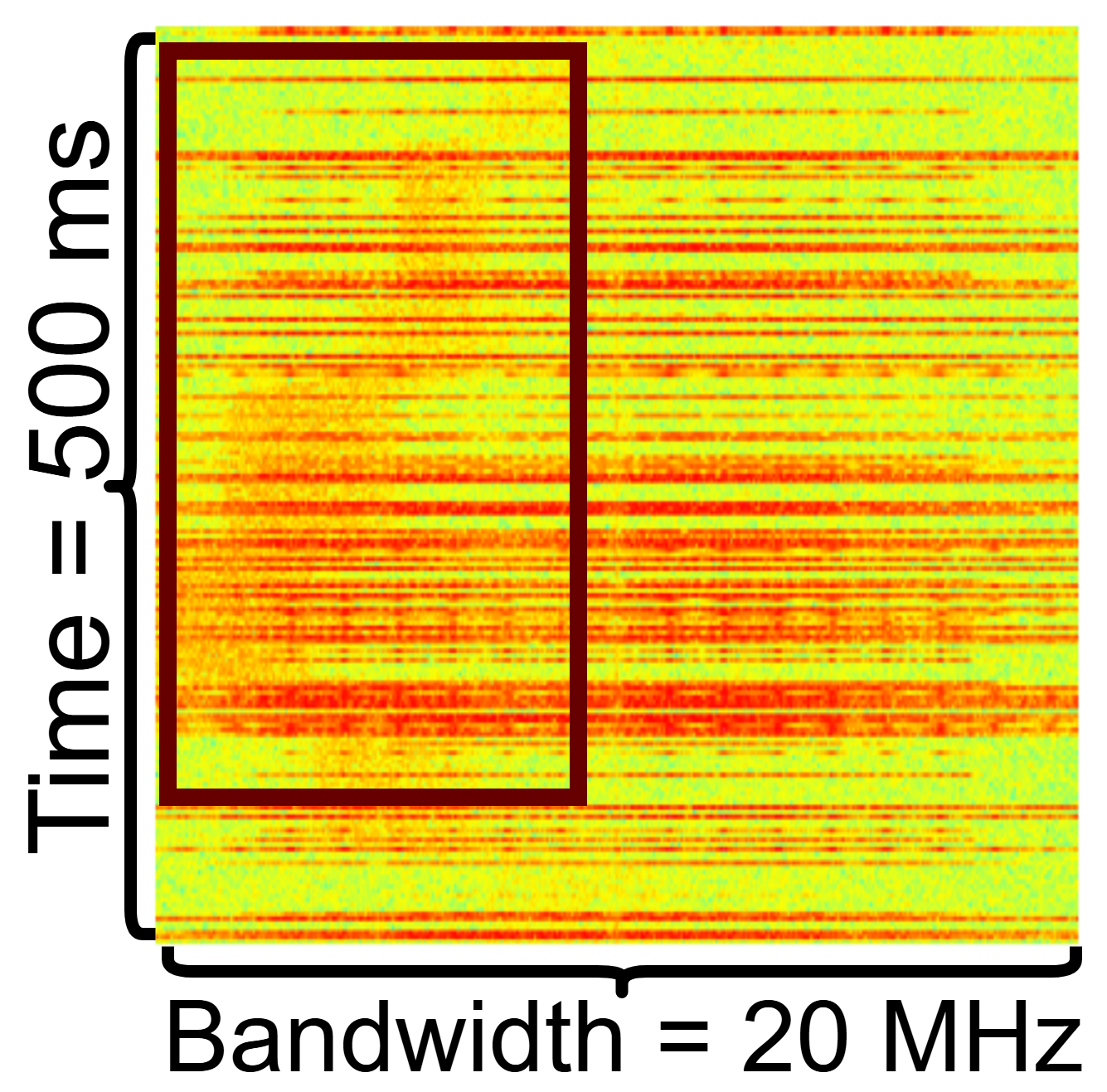}
        \caption{Wideband.}
        \label{fig:spec_wideband}
    \end{subfigure}

\caption{Collected over-the-air spectrograms illustrating benign transmissions and three distinct jamming attack types.}

    \label{fig:spectrograms_only}
\end{figure}
\noindent \textbf{Jamming Implementation.} Over-the-air interference between the edge devices and the central server was introduced by a single USRP X310 SDR, as shown in Fig.~\ref{fig:jammer}. To enable flexible and controlled interference without causing complete denial of service, we developed a suite of jamming scenarios implemented in Python using the UHD driver, allowing precise control over frequency, transmit power, activation timing, and spectral occupancy. We considered three jamming scenarios, primarily affecting data channels:
(i) \textit{Single-tone jamming} — continuous narrowband interference targeting a specific subset of subcarriers (12–36 subcarriers);
(ii) \textit{Pulsed jamming} — periodic bursts of narrowband interference over a broader subcarrier range (48-96 subcarriers), with configurable duty cycles and pulse durations (5–20 ms); and (iii) \textit{Wideband noise jamming} — wideband interference generated using Additive White Gaussian Noise (AWGN), spanning a configurable portion of the channel bandwidth (1, 2, 4, or 8 MHz), impacting a larger number of subcarriers simultaneously. Representative spectrograms corresponding to benign and jamming scenarios are illustrated in Fig.~\ref{fig:spectrograms_only}.

\noindent \textbf{Physical Deployment.} Lastly, Fig.~\ref{fig:lab_view} depicts the spatial arrangement of the wireless testbed components whose positions remained fixed throughout all experiments. The deployment spans a \(7 \times 10\,\mathrm{m}^2\) indoor area, with edge devices spaced at 1-meter intervals to capture spatial variability.

\begin{figure}[t]
    \centering
    \begin{subfigure}[b]{0.21\columnwidth}
        \captionsetup{justification=centering, labelformat=empty}
        \includegraphics[width=\linewidth]{images/spectrograms/spectrogram_single_tone.png}
        \caption{\hspace{-0.2em} (a) Tones.}
    \end{subfigure}
    \hfill
    \begin{subfigure}[b]{0.25\columnwidth}
        \captionsetup{justification=centering, labelformat=empty}
        \includegraphics[width=\linewidth]{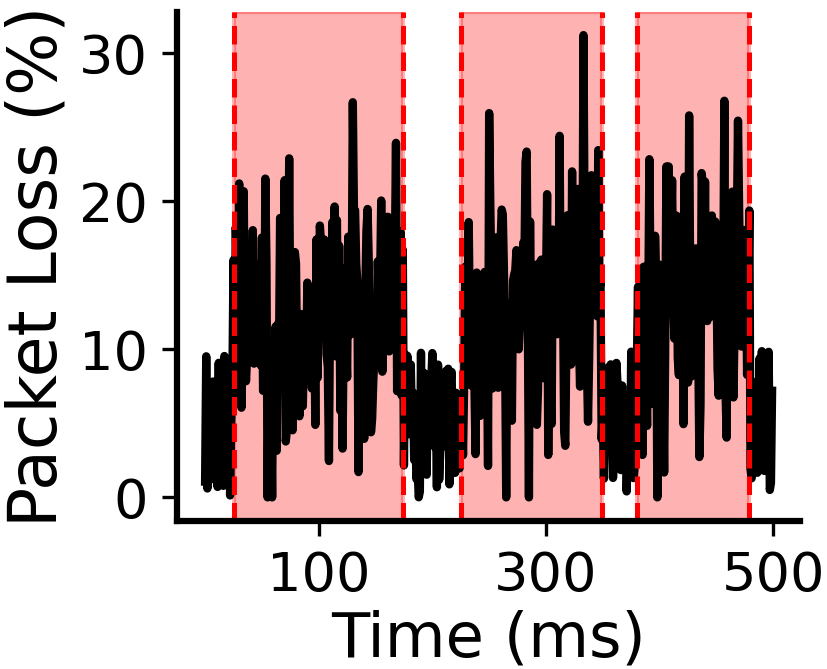}
        \caption{\hspace{-0.3em} (b) Packet loss.}
        \label{fig:kpi_packet_loss}
    \end{subfigure}
    \hfill
    \begin{subfigure}[b]{0.25\columnwidth}
        \captionsetup{justification=centering, labelformat=empty}
        \includegraphics[width=\linewidth]{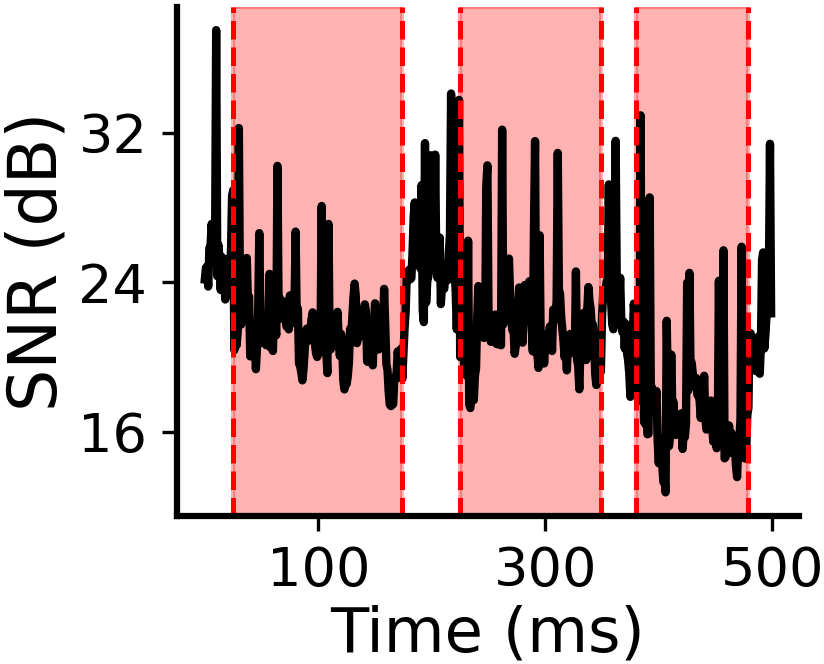}
        \caption{\hspace{-0.6em} (c) DL SNR.}
        \label{fig:kpi_snr}
    \end{subfigure}
    \hfill
    \begin{subfigure}[b]{0.25\columnwidth}
        \captionsetup{justification=centering, labelformat=empty}
        \includegraphics[width=\linewidth]{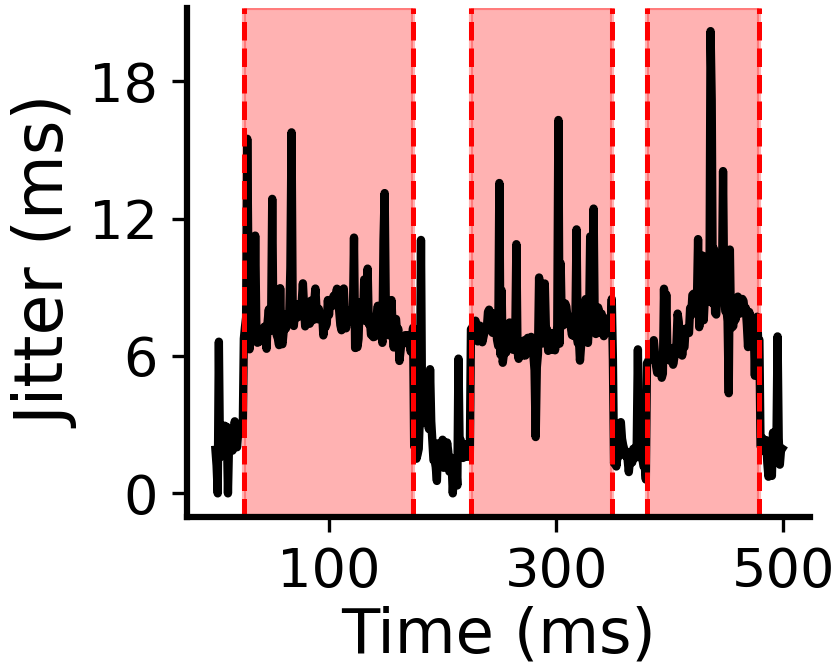}
        \caption{\hspace{-0.9em} (d) Jitter.}
        \label{fig:kpi_jitter}
    \end{subfigure}

    \begin{subfigure}[b]{0.21\columnwidth}
        \includegraphics[width=\linewidth]{images/spectrograms/spectrogram_pulse.png}
        \caption{Pulses.}
    \end{subfigure}
    \hfill
    \begin{subfigure}[b]{0.25\columnwidth}
        \includegraphics[width=\linewidth]{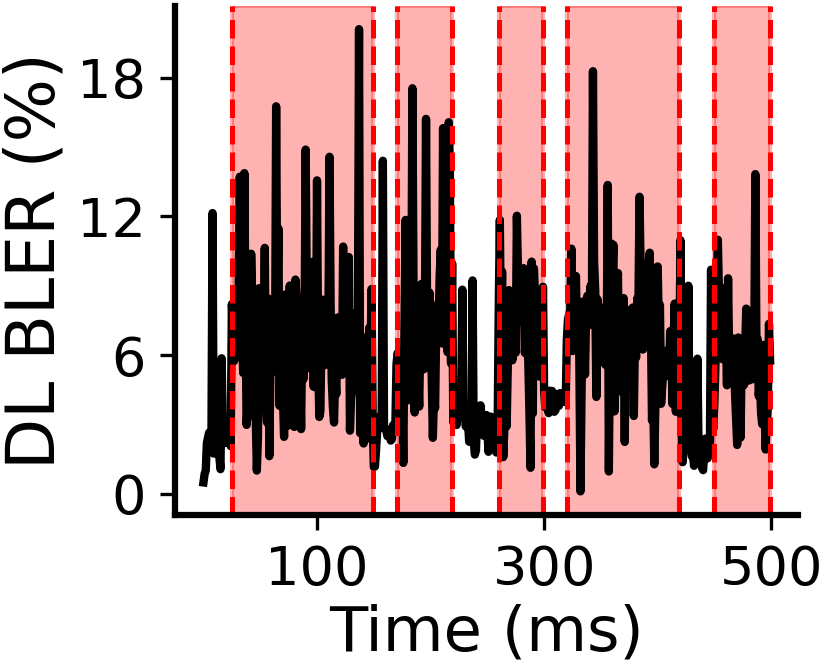}
        \caption{DL BLER.}
        \label{fig:bler}
    \end{subfigure}
    \hfill
    \begin{subfigure}[b]{0.25\columnwidth}
        \includegraphics[width=\linewidth]{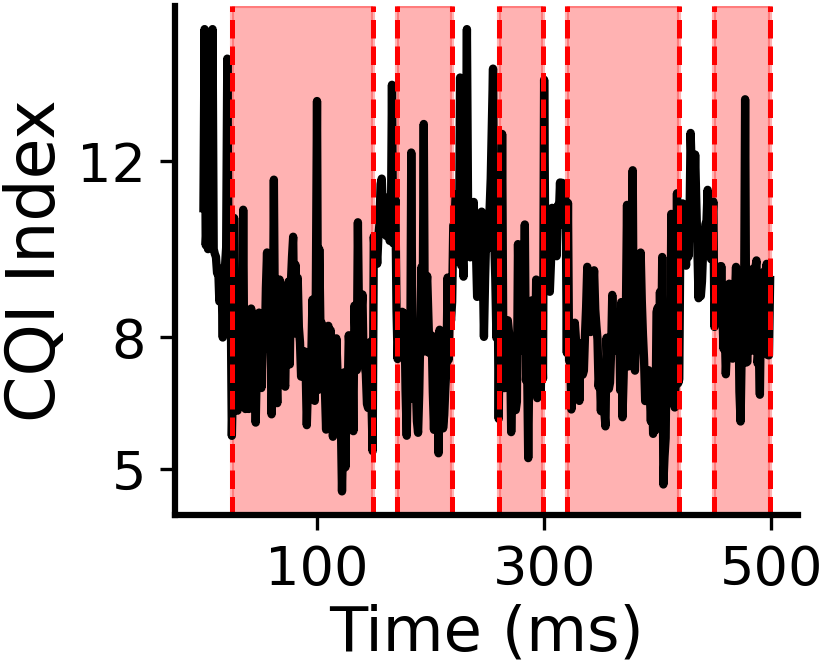}
        \caption{CQI.}
        \label{fig:cqi}
    \end{subfigure}
    \hfill
    \begin{subfigure}[b]{0.25\columnwidth}
        \includegraphics[width=\linewidth]{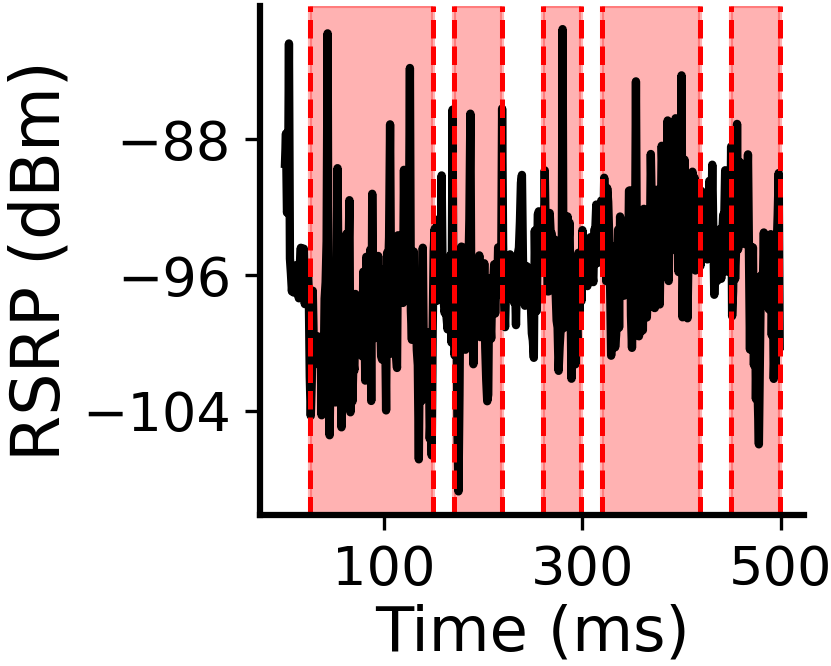}
        \caption{RSRP.}
        \label{fig:rsrp}
    \end{subfigure}
    \caption{Collected multimodal traces from an edge device, including time-aligned spectrogram images and KPI time-series for Wi-Fi (top row) and 5G (bottom row). Periods of jamming activity are highlighted with pink boxes.}
    \label{fig:kpi_multimodal}
\end{figure}

\subsection{Data Collection}

Due to the absence of publicly available multimodal datasets for jamming detection and classification, we utilized our testbed to jointly collect time-synchronized spectral data (as spectrogram images) and multi-channel network KPIs (as time series) for both Wi-Fi and 5G. Representative multimodal traces from our collected dataset are illustrated in Fig.~\ref{fig:kpi_multimodal}.

\noindent \textbf{Data Collection Pipeline.}
To enable fine-grained telemetry of spectral activity and network behavior, we implemented a fully configurable data collection pipeline on each Jetson device, supporting variable time granularities and adapting to diverse hardware and application constraints. For our experiments, we fixed the measurement window to 500 ms to balance temporal resolution with resource limitations. Within each 500 ms window, every device simultaneously captures two synchronized modalities: (1) a single spectrogram image representing spectral activity, and (2) a multi-channel time series of network KPIs. Spectrograms are recorded as 224$\times$224$\times$3 RGB images using the co-located USRP, which is dynamically tuned to the appropriate frequency band based on the device’s active network connection. Spectral snapshots are generated using a waterfall Fast Fourier Transform (FFT) representation with varying resolutions (256, 512, and 1024) to capture different levels of frequency granularity.  In parallel with spectrum activity capture, the device samples cross-layer KPIs from the wireless stack. For 5G-connected devices, KPIs are sampled at a 1 ms rate directly from the modem, while for Wi-Fi, the sampling rate is set to 5 ms due to system constraints, including driver and OS-level limitations. To address irregular sampling intervals in the collected KPIs we apply linear interpolation within each window to construct temporally aligned, multi-channel KPI sequences. Following this window-based collection approach over an extended period, each device independently collects and stores its multimodal data locally.

\begin{table}[t]
\centering
\caption{\textbf{Summary of experimental scenarios and sample distribution per wireless standard.}}
\label{tab:scenario_summary}
\begin{tabular}{@{}l@{\hspace{0.6em}}c@{\hspace{0.6em}}c@{\hspace{0.6em}}c@{}}
\toprule
\textbf{Category}     & \textbf{Gains (dB)} & \textbf{FFTs} & \textbf{Samples} \\ \midrule
Benign               & --          & 256, 512, 1024   & 9K (3K per FFT) \\
Single-Tone Jamming  & 10, 20, 30  & 256, 512, 1024   & 9K (1K per gain/FFT) \\
Pulsed Jamming       & 10, 20, 30  & 256, 512, 1024   & 9K (1K per gain/FFT) \\
Wideband Jamming     & 10, 20, 30  & 256, 512, 1024   & 9K (1K per gain/FFT) \\ \midrule
\textbf{Total}       & --          & --               & \textbf{36K} \\
\bottomrule
\end{tabular}
\end{table}

\noindent \textbf{Wi-Fi KPIs.}
For each device connected to Wi-Fi, we directly collected cross-layer KPIs and statistics from the Network Interface Card (NIC). Specifically, we collected the following metrics: \textit{TX failures}, \textit{TX packets retry exhausted}, \textit{TX total packets}, \textit{TX total packets sent}, \textit{TX data packets retried}, \textit{TX packets retries}, \textit{TX total bytes}, \textit{RX data packets}, \textit{RX total packets retried}, \textit{RX data bytes}, \textit{RSSI}, \textit{noise floor}, \textit{downlink SNR}, \textit{rate of last TX packet (min)}, \textit{rate of last TX packet (max)}, \textit{RTT to central server}, \textit{jitter}, and \textit{packet loss}.  
Note that while PHY and MAC-layer data were obtained directly from the NIC, higher-layer metrics such as \textit{RTT}, \textit{jitter}, and \textit{packet loss} were measured using custom-built UDP-based probing scripts between each device and the central server.

\noindent \textbf{5G KPIs.}
For each device connected to our 5G network, we extracted measurable KPIs directly from the 5G modem using custom parsing scripts. The collected metrics include: \textit{RSRP}, \textit{downlink SNR}, \textit{downlink BLER}, \textit{downlink MCS}, \textit{CQI}, and \textit{uplink buffer size}.

\noindent \textbf{Experimental Scenarios.}
Finally, we collected data across 30 scenarios for each wireless standard (Wi-Fi and 5G), capturing a diverse range of wireless conditions. This included 27 interference scenarios, spanning three types of jamming attacks (single-tone, pulsed, and wideband), each tested at three gain levels (10, 20, and 30 dB) and under three different receiver FFT resolutions (256, 512, and 1024), reflecting different device configurations and computational trade-offs. In addition, we conducted three scenarios under normal operating conditions without any intentional interference. Across all experiments, the physical locations of the devices and the jammer remained fixed, as shown in Fig.~\ref{fig:lab_view}.  The resulting dataset consists of 36{,}000 multimodal samples per wireless standard, as illustrated in Table~\ref{tab:scenario_summary}, uniformly distributed across all scenarios and devices.

\section{Evaluation Results}
\label{sec:evaluation}

\subsection{Baselines}

We compare \textit{FedJam} against state-of-the-art unimodal detection models, as well as a custom-implemented multimodal baseline, since, to the best of our knowledge, no prior multimodal jamming detection methods exist. More specifically, we evaluate \textit{FedJam} against the following baselines:

\begin{itemize}

    \item \textbf{FewShotDense}: A unimodal spectrogram-based convolutional neural network built on DenseNet121, previously used for few-shot interference classification tasks~\cite{federated_learning_spectrogram_5}.

    \item \textbf{SpectroNet}: A unimodal EfficientNet-B0-based architecture trained on spectrogram data for multi-class jamming detection and classification~\cite{spectrogram_1}.

    \item \textbf{JamShield}: A unimodal LSTM-based architecture trained on cross-layer KPI time series data for multi-class jamming detection~\cite{jamshield}.

    \item \textbf{SpectroKPI-Fuser}: A customized EfficientNet-B0 convolutional model that performs early fusion by stacking synchronized $N$-channel KPI time series with 3-channel spectrograms into a unified $(N{+}3)$-channel input.

    \item \textbf{Notable Unimodal Models}: We also experiment with two lightweight unimodal architectures to assess individual modality performance under constrained resources: (i) an \textit{1D-CNN}~\cite{kiranyaz20211d} operating on KPI time series and (ii) \textit{TinyViT}~\cite{wu2022tinyvit}, a compact vision transformer applied to spectrogram inputs.

\end{itemize}

\noindent For all baselines, we adopted the original training configurations as specified by the respective authors.

\subsection{FedJam Configuration}

For spectral feature extraction, the \textit{Spectrum Encoder} employs \textit{MobileNetV3-Small}~\cite{howard2019searching}, a lightweight convolutional neural network pretrained on ImageNet, producing a 1024-dimensional feature vector that captures localized interference patterns in the frequency domain. The \textit{KPI Encoder} is implemented using a compact Transformer-based architecture composed of four self-attention layers, preceded by a linear projection to 128 dimensions. The resulting sequence is aggregated using an attention-based pooling mechanism to produce a 128-dimensional embedding summarizing the network’s temporal performance. The \textit{Fusion Module} combines the two modality-specific embeddings via concatenation, forming a 1152-dimensional joint representation. This fused vector is then passed to the \textit{Multimodal Projection Head Module}, implemented as a two-layer fully connected neural network with 256 hidden units and ReLU activation, followed by an output layer producing logits over the four jamming classes. Finally, all of \textit{FedJam}'s components are jointly fine-tuned end-to-end using our collected multimodal dataset.

\subsection{Training Details}

The dataset was partitioned into training and testing subsets using an 80\%-20\% split. Data augmentation was applied to improve model robustness, including normalization with channel-wise means of [0.485, 0.456, 0.406] and standard deviations of [0.229, 0.224, 0.225]. Additional augmentations included random brightness adjustment, flips, rotations, color jitter, salt-and-pepper noise, Gaussian noise, and cutout, implemented via a custom augmentation pipeline. Input images were resized to $224 \times 224$ pixels. 
KPI time-series were normalized per channel using min-max scaling based on values collected across both benign and interference scenarios. The KPI sequences were downsampled to 256 time steps per window through the ATD, using fixed-stride sampling. 

Model training used the Adam optimizer with a cosine annealing learning rate schedule, starting at $5 \times 10^{-4}$ and decaying to $1 \times 10^{-6}$ after a warmup period. We used a batch size of 128 and one local epoch per federated averaging round, with all clients participating in each round. Cross-entropy loss served as the objective function. We employ the standard \textit{FedAvg} algorithm for aggregation, unless otherwise stated. Detection accuracy is evaluated on a held-out test set at the server side after each round. All reported results are averaged over three random seeds.

\subsection{Evaluation Results}

We evaluate \textit{FedJam} in two complementary settings: (i) a simulation-based evaluation to benchmark detection accuracy, convergence behavior, and scalability under varying data distributions, number of participating devices, and configurations; and (ii) a real-world deployment to quantify system-level performance, including training latency, inference latency, and resource overhead. While our data collection spans both Wi-Fi and 5G, we focus on the Wi-Fi dataset, which offers a broader and more diverse set of cross-layer KPIs.

\noindent\textbf{Simulation Results}

\noindent For our simulation experiments, we utilized the \textit{Flower} framework \cite{beutel2020flower}, a widely adopted and flexible platform in the FL community that enables evaluation of federated solutions across diverse scales and heterogeneous settings.

\begin{figure}[t]
  \centering
  \begin{subfigure}{0.48\linewidth}
    \centering
    \captionsetup{justification=centering, labelformat=empty}
    \includegraphics[width=\linewidth]{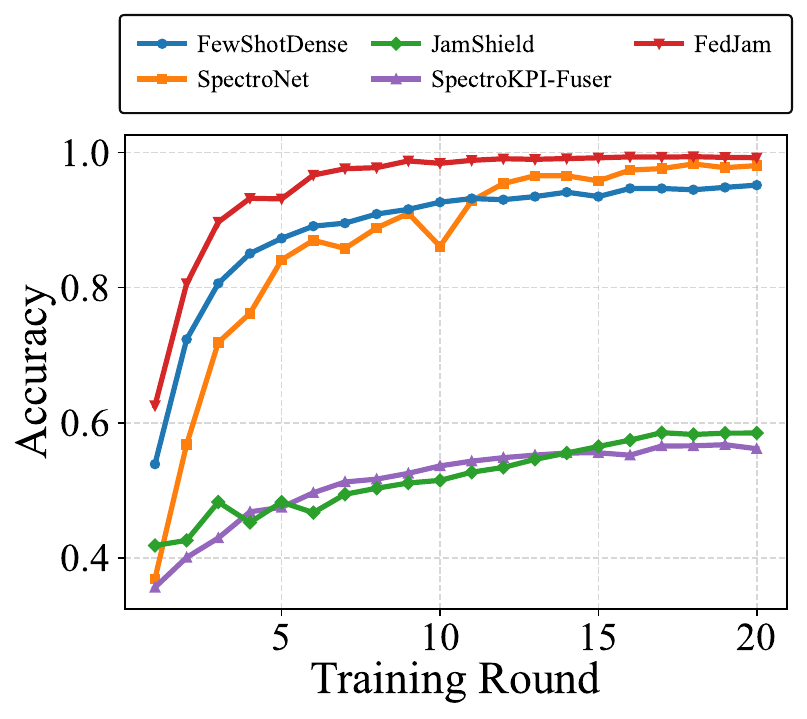}
    \caption{\hspace{1.5em} (a) IID setting (10 clients).}
    \label{fig:accuracy_plot_iid}
  \end{subfigure}
  \hspace{-0.015\linewidth}
  \begin{subfigure}{0.48\linewidth}
    \centering
    \captionsetup{justification=centering, labelformat=empty}
    \includegraphics[width=\linewidth]{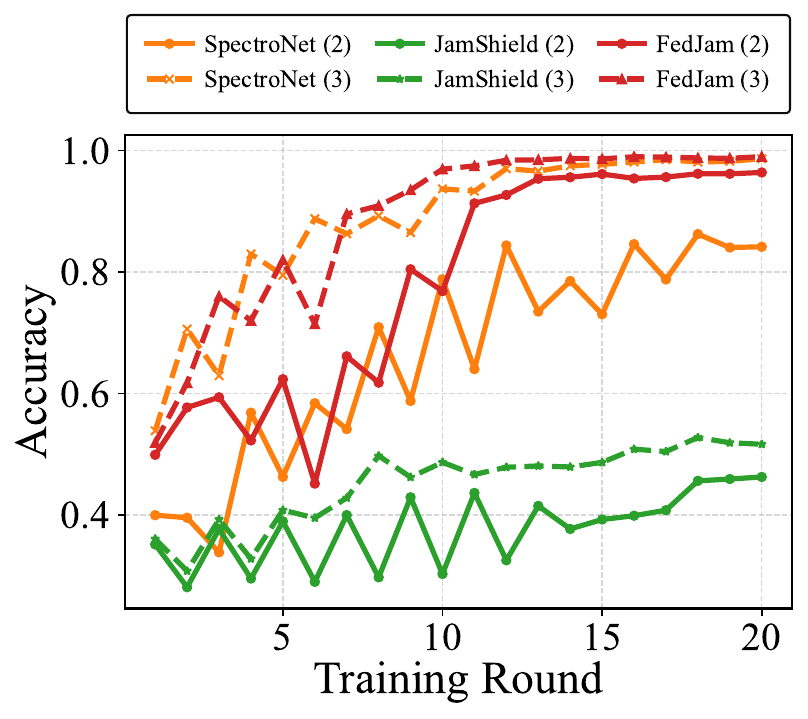}
    \caption{\hspace{0.1em} (b) Non-IID setting (10 clients).}
    \label{fig:accuracy_plot_niid}
  \end{subfigure}
  \caption{Comparison of detection accuracy across IID and Non-IID training settings.}
  \label{fig:accuracy_comparison}
\end{figure}

\noindent\underline{\textit{1)  Performance on IID Settings.}} Table~\ref{tab:accuracy_pivot} presents the detection accuracy at convergence for various baselines and client configurations, under an IID setting, where each client receives the same number of samples for each of the four classes. \textit{FedJam} consistently outperforms all baselines, achieving $99.2\%$ accuracy with 10 clients and maintaining strong performance ($92.6\%$) even in the more distributed 50-client scenario. Notably, KPI-focused baselines such as 1D-CNN and JamShield exhibit consistently poor accuracy, particularly as the number of clients increases, indicating that they are not well-suited for the jamming detection task on their own. Interestingly, while TinyViT, given its attention-based approach, is considered state-of-the-art on general image recognition tasks, it underperforms here, likely due to the relatively structured and domain-specific nature of spectrogram images. This suggests that having more parameters does not necessarily translate to better performance for jamming detection. The detection accuracy curve in Figure~\ref{fig:accuracy_plot_iid} further demonstrates the rapid convergence of \textit{FedJam}, which reaches 90\% accuracy \textit{within just 3 communication rounds}, more than twice as fast as the next best method, which requires 8 rounds to reach the same threshold. This fast adaptation can be particularly valuable in online or resource-constrained learning settings, where rapid learning from limited rounds is crucial. 

\input{table_iid}

\noindent\underline{\textit{2) Performance on Non-IID Settings.}} In real-world jamming scenarios, edge devices (e.g., radios or IoT nodes) are often exposed to only a subset of jamming patterns due to their location, environment, and channel conditions. Table~\ref{tab:non_iid_grouped} presents the detection accuracy at convergence under increasingly challenging Non-IID settings, where each client only observes a subset of the total four classes during training. Specifically, the “2” and “3” columns correspond to cases where each client has access to only two or three of the four classes, respectively. \textit{FedJam} demonstrates significant resilience and performance superiority in both Non-IID settings, especially under the more extreme Non-IID-2 condition. For instance, with 10 clients and only 2 classes per client, \textit{FedJam} achieves $96.38\%$ accuracy, substantially outperforming the next-best baseline, SpectroNet, which reaches $84.15\%$. Similar trends hold across larger federations (25 and 50 clients), where FedJam maintains high accuracy ($91.28\%$ and $86.58\%$, respectively), while baseline methods suffer sharper drops. The convergence behavior, shown in Figure~\ref{fig:accuracy_plot_niid}, further illustrates this advantage: \textit{FedJam} not only achieves higher final accuracy but also converges faster than SpectroNet, the strongest unimodal baseline. These results show \textit{FedJam} can fuse diverse modalities and generalize well despite imbalanced or incomplete data, making it fit for real-world, uneven client settings.

\input{table_non_iid}

\begin{figure}[t]
  \centering
  \begin{subfigure}[b]{0.49\columnwidth}
    \centering
    \captionsetup{justification=centering, labelformat=empty}
    \includegraphics[width=\linewidth]{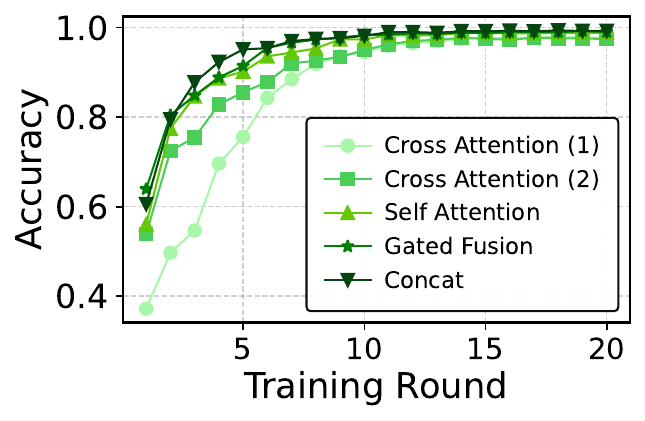}
    \caption{\hspace{1.5em} (a) Fusion (10 clients).}
    \label{fig:fusion}
  \end{subfigure}
  \hfill
  \begin{subfigure}[b]{0.49\columnwidth}
    \centering
    \captionsetup{justification=centering, labelformat=empty}
    \includegraphics[width=\linewidth]{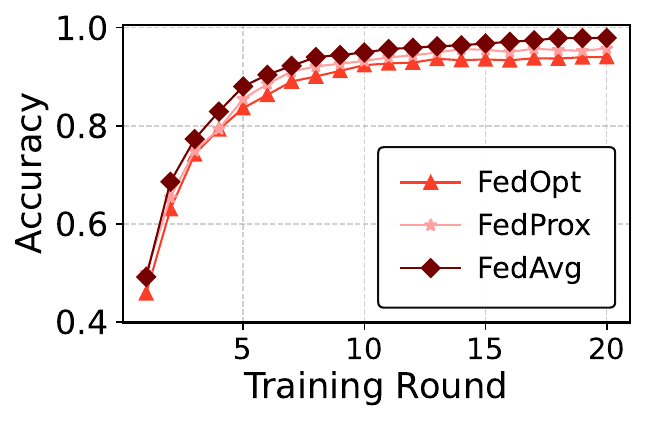}
    \caption{\hspace{1.5em} (b) Aggregation (25 clients).}
    \label{fig:fed_algo}
  \end{subfigure}
  \caption{Impact of (a) fusion design and (b) federated optimization strategy on detection performance.}
  \label{fig:fusion_fed}
\end{figure}

\noindent\underline{\textit{3) Fusion \& Aggregation.}} An essential component of any multimodal framework is the choice of fusion mechanism, which determines how information from different modalities is combined. To identify the most effective approach, we compare four representative fusion strategies: \textit{concatenation}, \textit{cross-attention}, \textit{self-attention}, and \textit{gated fusion}. For cross-attention, we explore two variants: images attending to time-series features (1), and vice versa (2). These methods span a spectrum of complexity and capacity to model inter-modal interactions. As shown in Fig.~\ref{fig:fusion}, although all methods achieve comparable performance, simple concatenation slightly outperforms the others. Given its low computational overhead and stable accuracy, we adopt concatenation in our final design to balance efficiency and effectiveness.

In addition to fusion strategies, the choice of aggregation algorithm can influence convergence and robustness in FL. We evaluate three widely used aggregation methods: \textit{FedAvg}, \textit{FedOpt}~\cite{fedopt}, and \textit{FedProx}~\cite{fedprox}, to assess their compatibility with our framework. As shown in Fig.~\ref{fig:fed_algo}, all algorithms perform similarly, with \textit{FedAvg} offering a marginal improvement. These results highlight that our method does not rely on a specific aggregation scheme. Indeed, \textit{FedJam} is designed to be modular and aggregation-agnostic, enabling seamless integration with future or domain-specific optimizers, and supporting flexible experimentation in practical deployments.

\noindent \underline {\textit{3) Effect of FFT Size and Jamming Gain on Model Accuracy.}} To understand the impact of spectrogram resolution and jamming intensity on detection performance, we explore two controlled dataset variations: frequency resolution (FFT size), which reflects the spectrogram granularity supported by the clients, and jamming signal strength (gain).

\begin{figure}[t]
  \centering
  \begin{subfigure}[b]{0.49\columnwidth}
    \centering
    \captionsetup{justification=centering, labelformat=empty}
    \includegraphics[width=\linewidth]{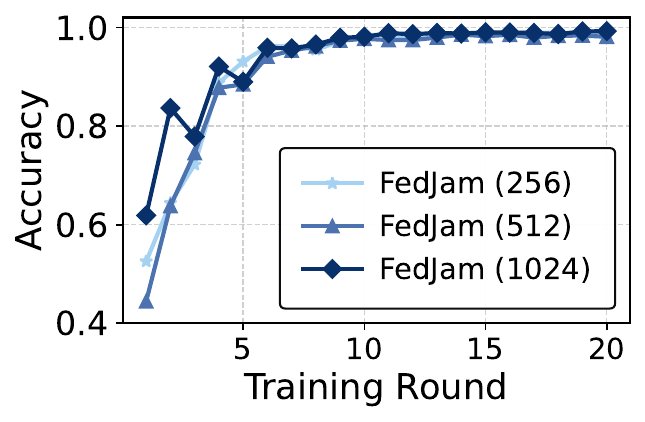}
    \caption{\hspace{1.5em} (a) Varying FFT resolution.}
    \label{fig:fft_eval}
  \end{subfigure}
  \hfill
  \begin{subfigure}[b]{0.49\columnwidth}
    \centering
    \captionsetup{justification=centering, labelformat=empty}
    \includegraphics[width=\linewidth]{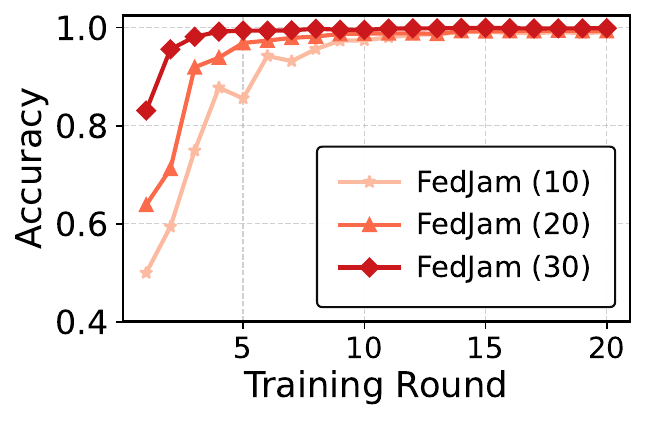}
    \caption{\hspace{1.5em} (b) Jammer gain level (dB).}
    \label{fig:gain_eval}
  \end{subfigure}
  \caption{Impact of FFT size and jammer gain on detection.}
  \label{fig:fft_gain_eval}
\end{figure}

\textbf{FFT Variations.} We create three variants of the dataset using spectrograms computed with FFT sizes of 256, 512, and 1024, respectively. These correspond to increasing levels of frequency resolution. As shown in Fig.~\ref{fig:fft_eval}, using 1024-point FFTs leads to marginally better performance across all training rounds, confirming that higher-resolution frequency content provides slightly stronger cues for jamming detection. However, the gains are modest—indicating that models trained with lower-resolution inputs still perform competitively. This suggests that in resource-constrained environments, users can safely reduce FFT size to lower compute and memory overhead without significantly sacrificing accuracy.

\textbf{Gain Variations.} We next analyze how the strength of jamming signals affects detection performance. To do so, we generate three dataset versions corresponding to jamming gains of 10, 20, and 30 dB. As shown in Fig.~\ref{fig:gain_eval}, the model detects higher-gain attacks more reliably and converges faster. For instance, with 30 dB gain, the model reaches nearly 95\% accuracy within just two training rounds. This aligns with intuition: stronger jamming signals have more distinguishable  footprints, making them easier to learn and classify. The results validate the robustness of \textit{FedJam} across a range of interference intensities and demonstrate its capacity to generalize across different interference conditions.

\begin{figure}[t]
    \centering
    \includegraphics[width=0.835\linewidth]{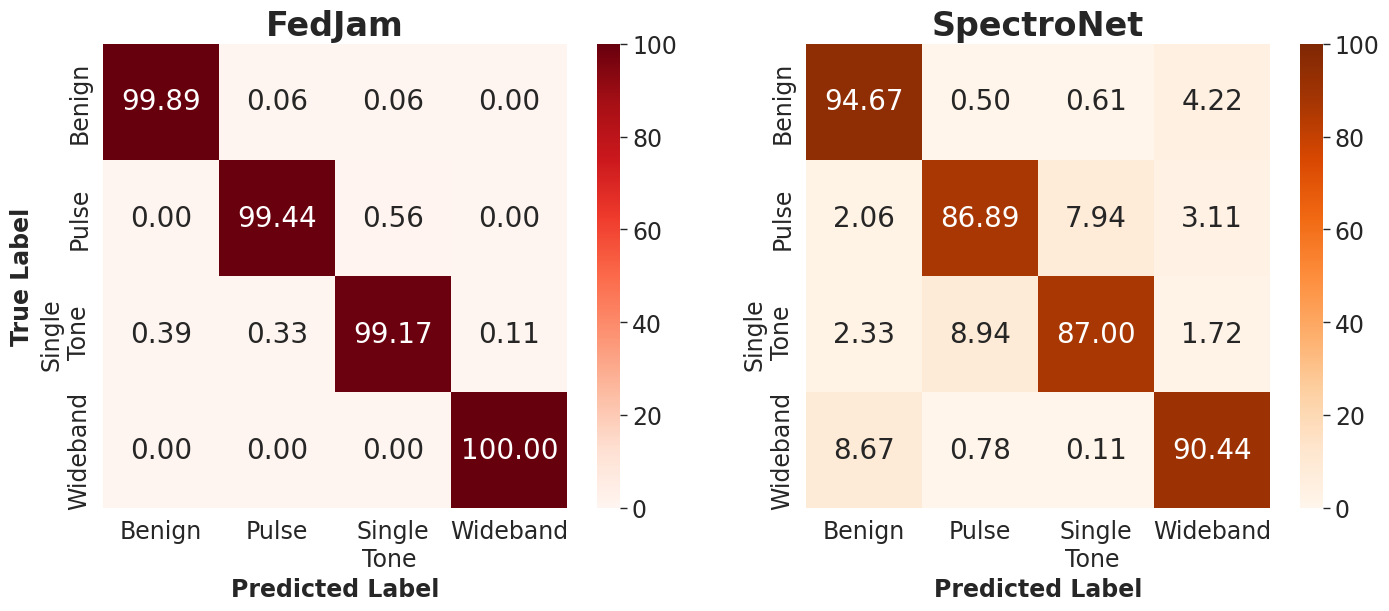}
    \caption{Normalized confusion matrices for FedJam (left) and SpectroNet (right) across all classes.}
    \label{fig:confusion_matrix}
\end{figure}

\noindent \underline {\textit{4) Per-Class Performance Breakdown.}} To gain a deeper understanding of per-class performance, we visualize the normalized confusion matrices for our proposed \textit{FedJam} model and the best-performing baseline, \textit{SpectroNet}, as shown in Fig.~\ref{fig:confusion_matrix}. These matrices reveal how well each model distinguishes between different jamming types and the benign class.

Both models achieve high accuracy on the \textit{benign} class, reflecting their ability to avoid false alarms in the absence of interference. However, key differences emerge in the jamming classes. In particular, \textit{SpectroNet} struggles to distinguish between the \textit{pulse} and \textit{single-tone} jamming classes, with classification accuracies dropping below 90\%, given their similar structures, as shown in Figure~\ref{fig:spectrograms_only}. This confusion suggests that SpectroNet's feature representations are insufficiently discriminative for attacks with similar spectral profiles. In contrast, \textit{FedJam} exhibits more consistent and reliable performance across all classes, highlighting the strength of its multimodal fusion in learning nuanced signal patterns. This insight reinforces the value of our multimodal approach, especially in scenarios where distinguishing between fine-grained jamming types is critical for appropriate mitigation and response

\noindent\textbf{Real-World Testbed Results} 

\begin{figure}[t]
  \centering

  \begin{subfigure}[b]{0.48\columnwidth}
    \centering
    \captionsetup{justification=centering, labelformat=empty}
    \includegraphics[width=\linewidth]{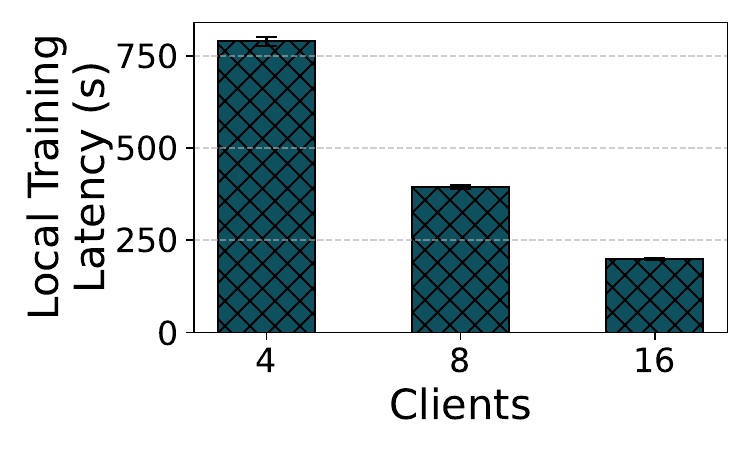}
    \caption{\hspace{2em} (a) Local training latency.}
    \label{fig:latency_training}
  \end{subfigure}
  \hfill
  \begin{subfigure}[b]{0.48\columnwidth}
    \centering
    \captionsetup{justification=centering, labelformat=empty}
    \includegraphics[width=\linewidth]{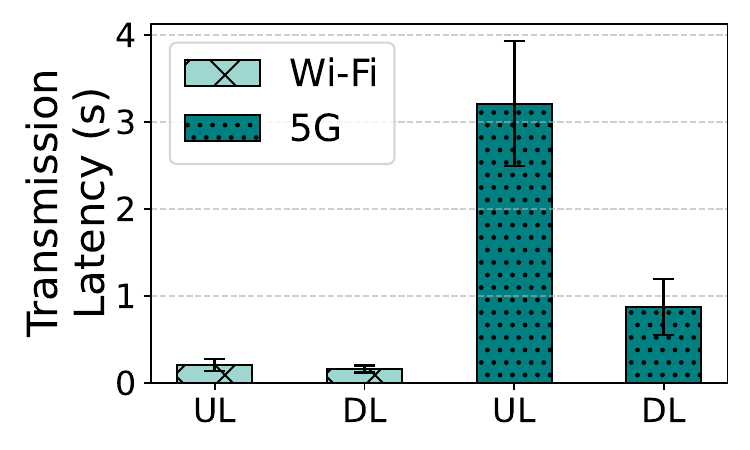}
    \caption{\hspace{1.5em} (b) Communication latency.}
    \label{fig:latency_transmission}
  \end{subfigure}

  \begin{subfigure}[b]{0.48\columnwidth}
    \centering
    \captionsetup{justification=centering, labelformat=empty}
    \includegraphics[width=\linewidth]{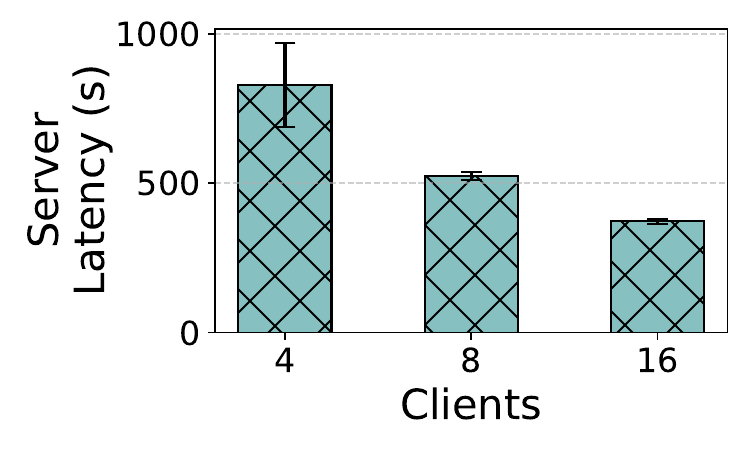}
    \caption{\hspace{2em} (c) Server latency.}
    \label{fig:latency_aggregation}
  \end{subfigure}
  \hfill
  \begin{subfigure}[b]{0.48\columnwidth}
    \centering
    \captionsetup{justification=centering, labelformat=empty}
    \includegraphics[width=\linewidth]{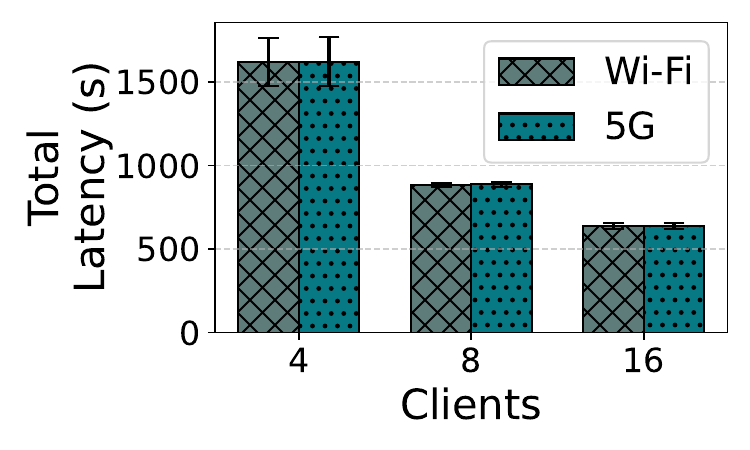}
    \caption{\hspace{2em} (d) Training round latency.}
    \label{fig:latency_total}
  \end{subfigure}

  \caption{Average training round latency decomposition across different numbers of clients and wireless standards.}
  \label{fig:latency_eval}
\end{figure}

\noindent \underline {\textit{1) End-to-End Federated Training Latency Analysis.}} We first analyze the per-round training latency for different number of participating devices and wireless backhauls. The total per-round latency (Fig.\ref{fig:latency_eval}d)) is decomposed into three components: (i) local training latency (Fig.\ref{fig:latency_eval}a), (ii) communication latency (Fig.\ref{fig:latency_eval}b), and (iii) server-side latency (Fig.\ref{fig:latency_eval}c).  As illustrated in Fig.~\ref{fig:latency_eval}a), the local training latency decreases with an increasing number of participating devices, as the dataset is partitioned across more devices, reducing the per-device computation load. Communication latency, reflecting both the downlink transmission of the global model to edge devices and the uplink transmission of locally trained weights, varies notably with the wireless backhaul. Specifically, 5G incurs substantially higher uplink delay than Wi-Fi, primarily due to limited uplink slot allocations in the radio frame structure and increased control-plane signaling overhead, as shown in Fig.~\ref{fig:latency_transmission}. In contrast, server-side latency, which includes the waiting time for model update reception from the clients, weight aggregation, and global model evaluation, remains relatively stable across configurations, as depicted in Fig.~\ref{fig:latency_aggregation}. Finally, as shown in Fig.\ref{fig:latency_total}, end-to-end latency remains comparable across Wi-Fi and 5G, as \textit{FedJam}’s lightweight architecture (12.52 MB/round; Table \ref{tab:model_sizes}) imposes minimal backhaul load and transmission delay.

\begin{table}[t]
    \centering
    \caption{\textbf{Model size (FP32) and parameter comparison.}}
    \begin{tabular}{lcc}
    \toprule
    \textbf{Model} & \textbf{Parameters (M)} & \textbf{Size (MB)} \\
    \midrule
    FewShotDense         & 7.98   & 32.0  \\
    SpectroNet           & 5.30   & 20.4  \\
    JamShield            & 0.34   & 1.28  \\
    SpectroKPI-Fuser     & 5.64   & 21.72  \\
    FedJam               & 3.28   & 12.52   \\
    \bottomrule
    \end{tabular}
    \label{tab:model_sizes}
\end{table}

\begin{figure}[t]
  \centering
  \includegraphics[width=0.7\columnwidth]{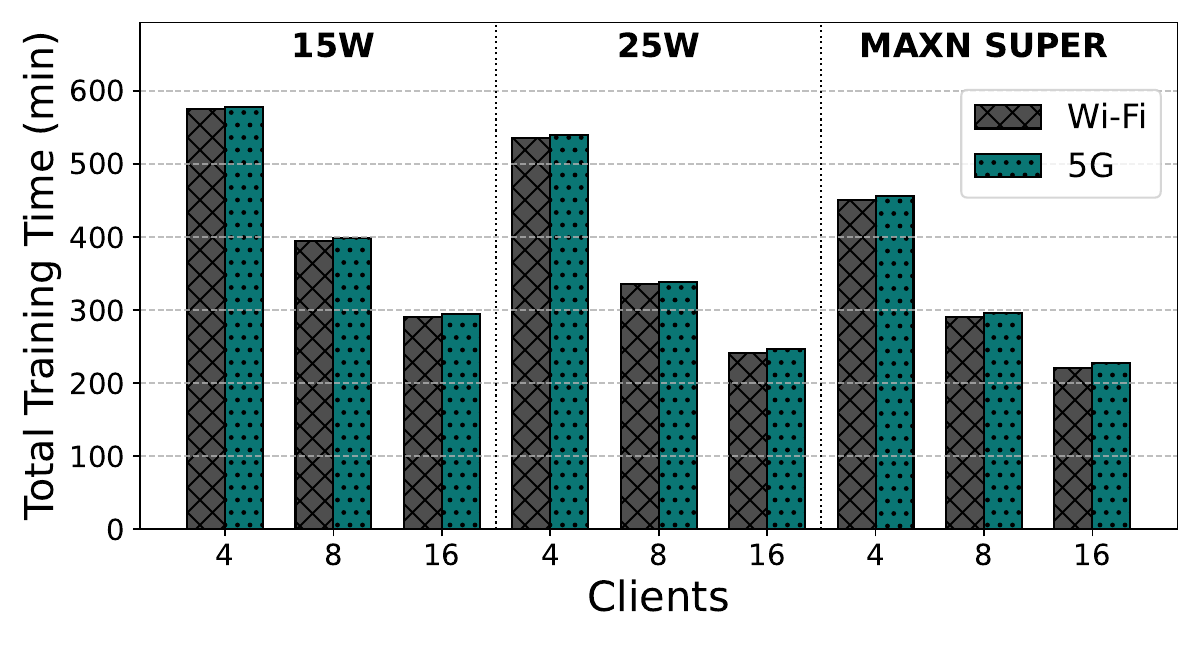}
  \caption{Total training time across different energy modes, number of participating clients and wireless backhauls.}

  \label{fig:energy_latency}
\end{figure}

We additionally evaluate the total training time required under different device energy modes, where all participating clients operate at the same power level. The Jetson Orin Nano devices support three energy modes: Mode 0 (15W), Mode 1 (25W), and Mode 2 (MAXN SUPER), which progressively scale CPU/GPU frequencies and memory bandwidth~\cite{energy_modes}. As shown in Fig.~\ref{fig:energy_latency}, higher energy modes consistently reduce training time across both Wi-Fi and 5G. For instance, under Wi-Fi with 4 clients, total training time decreases from 575 minutes (15W) to 451 minutes (MAXN SUPER), corresponding to a 21.6\% reduction. Similarly, with 16 clients, training time drops from 290 to 220 minutes—a 24.1\% improvement. These improvements reflect lower per-device training time due to increased hardware utilization and faster computation, although diminishing gains are observed beyond 8 clients due to reduced per-client workload.

\begin{figure}[t]
    \centering
    \begin{subfigure}{0.33\linewidth}
        \centering
        \captionsetup{justification=centering, labelformat=empty}
        \includegraphics[width=\linewidth]{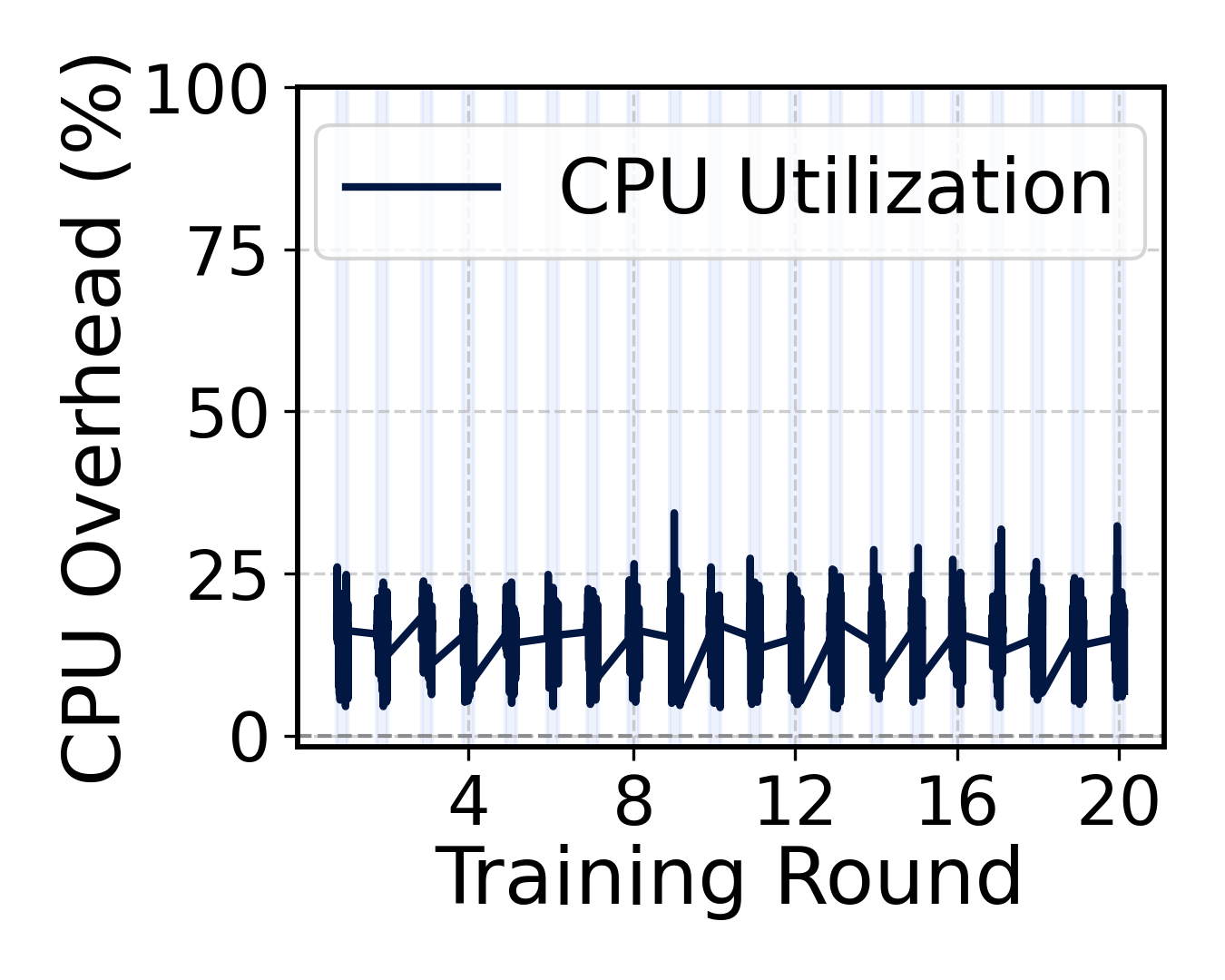}
        \caption{\hspace{0.1em} (a) CPU utilization.}
        \label{fig:cpu-overhead}
    \end{subfigure}
    \hspace{-0.8em}
    \begin{subfigure}{0.33\linewidth}
        \centering
        \captionsetup{justification=centering, labelformat=empty}
        \includegraphics[width=\linewidth]{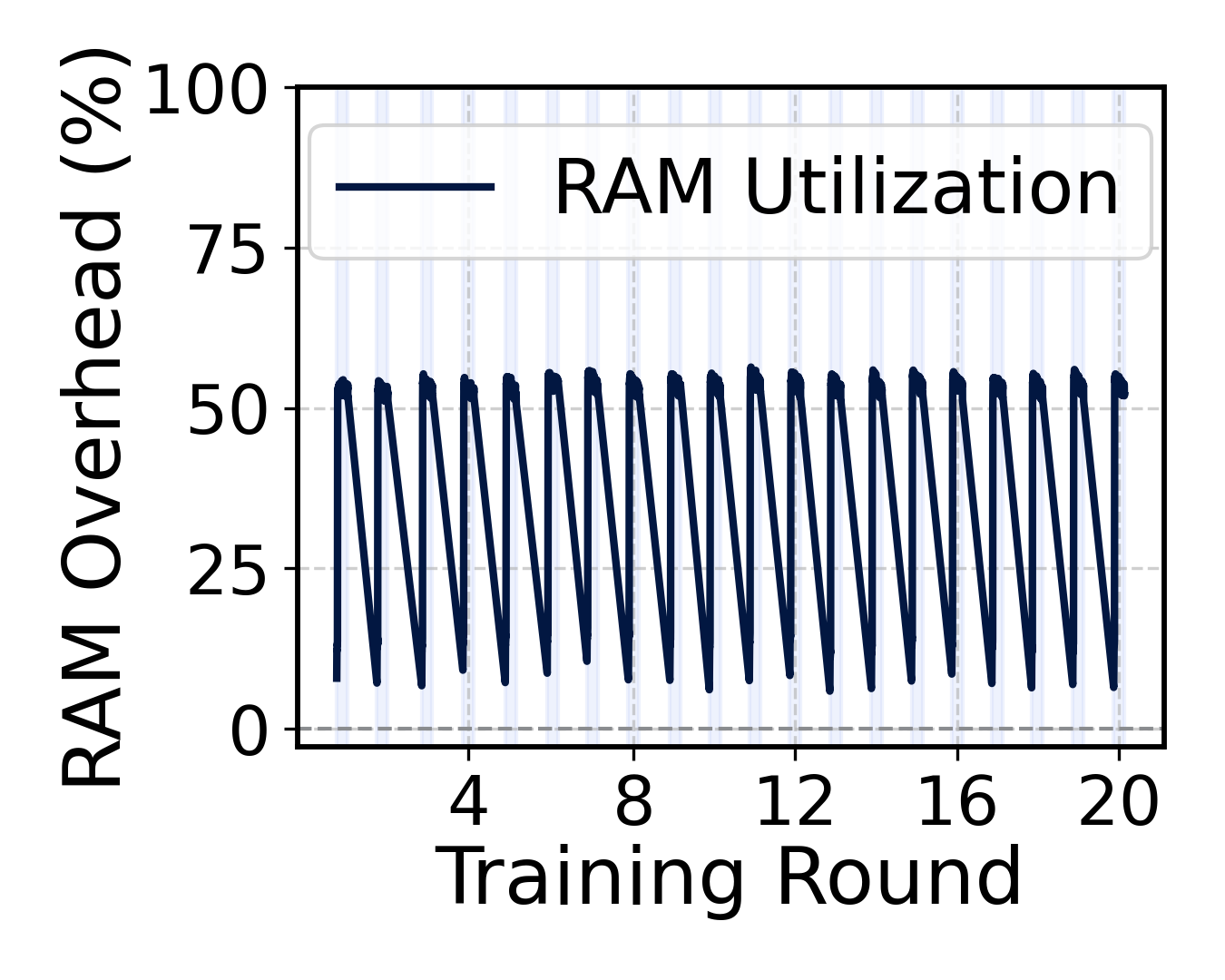}
        \caption{\hspace{0.1em} (b) RAM utilization.}
        \label{fig:ram-overhead}
    \end{subfigure}
    \hspace{-0.8em}
    \begin{subfigure}{0.33\linewidth}
        \centering
        \captionsetup{justification=centering, labelformat=empty}
        \includegraphics[width=\linewidth]{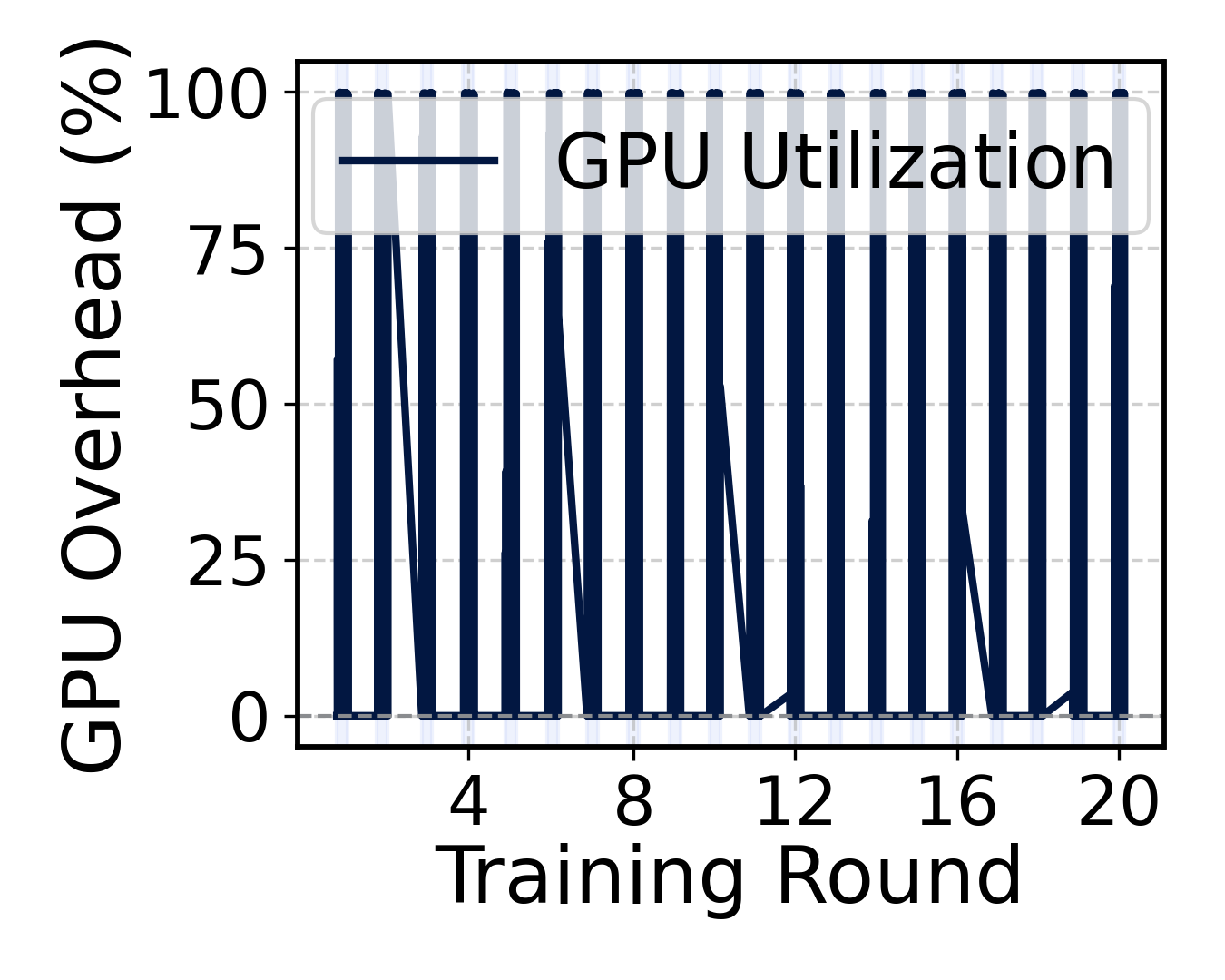}
        \caption{\hspace{0.1em} (c) GPU utilization.}
        \label{fig:gpu-overhead}
    \end{subfigure}

    \caption{Client resource overhead during training phase.}
    \label{fig:jetson-overheads}
\end{figure}

\begin{table}[t]
  \centering
  \caption{\textbf{Average client resource utilization across different energy modes during training phase.} Column “Clients” indicates the number of dataset partitions distributed across participating clients.}
  \begin{tabular}{@{}ccccc@{}}
    \toprule
    \textbf{Energy Mode} & \textbf{Clients} & \textbf{CPU (\%)} & \textbf{Memory (\%)} & \textbf{GPU (\%)} \\
    \midrule
    \multirow{3}{*}{15W} 
      & 4  & 12.4 ± 3.4 & 51.3 ± 8.9  & 6.6 ± 24.3 \\
      & 8  & 13.7 ± 3.3 & 50.6 ± 8.6  & 6.4 ± 24.1 \\
      & 16 & 13.2 ± 3.3 & 48.3 ± 8.4  & 6.4 ± 24.3 \\
    \midrule
    \multirow{3}{*}{25W} 
      & 4  & 16.3 ± 2.9 & 54.9 ± 5.9 & 5.7 ± 22.9 \\
      & 8  & 16.8 ± 2.8 & 46.8 ± 7.7 & 5.6 ± 22.8 \\
      & 16 & 16.7 ± 2.7 & 50.5 ± 9.9 & 5.7 ± 22.8 \\
    \midrule
    \multirow{3}{*}{MAXN SUPER} 
      & 4  & 14.3 ± 2.9 & 50.8 ± 9.2 & 6.8 ± 24.7 \\
      & 8  & 7.1 ± 3.3 & 47.6 ± 8.8 & 5.6 ± 22.5 \\
      & 16 & 6.8 ± 2.4 & 48.7 ± 8.7 & 5.8 ± 23.3 \\
    \bottomrule
  \end{tabular}
  
  \vspace{-1em}
  \label{tab:energy_utilization}
\end{table}

\noindent \underline{\textit{2) Client-Side Training Resource Overhead.}} We next profile the system-level resource footprint introduced solely by \textit{FedJam} on device-level during training. Fig.~\ref{fig:jetson-overheads} presents the computation and memory overhead observed during training runtime,  while Table~\ref{tab:energy_utilization} summarizes the average resource usage under different energy modes and number of clients. Across all configurations, \textit{FedJam} imposes minimal resource overhead. CPU usage remains below 17\%, memory usage stays within 45–55\%, and GPU utilization is consistently under 7\%. Notably, higher device energy modes lead to slight reductions in computation and memory overhead due to improved hardware throughput and faster task execution.

\noindent \underline{\textit{3) Total Classification Delay Breakdown.}} Finally, we measure the end-to-end time required by \textit{FedJam} to collect input data and generate classification decisions. As illustrated in Table~\ref{tab:inference_latency_breakdown}, the overall delay comprises three components: input collection (spectrogram and KPI sampling), preprocessing (e.g., synchronization, spectrogram resizing, and KPI normalization), and model inference. Inference accounts for less than 5\% of the total latency, with the majority attributed to input acquisition and preprocessing. The delay remains consistent across different device energy modes, as the inference pipeline is lightweight and not bottlenecked by compute constraints. Overall, the analysis confirms that \textit{FedJam} supports timely, low-latency inference suitable for on-device deployment in real-world resource-constrained environments.

\begin{table}[h]
  \centering
  \caption{\textbf{Breakdown of FedJam latency components.}}
  \begin{tabular}{@{}lc@{}}
    \toprule
    \textbf{Stage} & \textbf{Latency (ms)} \\
    \midrule
    Spectrogram + KPI Collection & 501.2 ± 10.2 \\
    Preprocessing                & 347.4 ± 15.6 \\
    Inference                    & 34.6 ± 6.7 \\
    \midrule
    \textbf{Total Latency}       & \textbf{884.6 ± 21.0} \\
    \bottomrule
  \end{tabular}
  \label{tab:inference_latency_breakdown}
\end{table}

\section{Conclusion}
\label{sec:conclusion}

In this paper, we introduced \textit{FedJam}, a multimodal FL framework for efficient jamming detection and classification. Unlike centralized approaches, \textit{FedJam} enables on-device training and inference on resource-constrained edge devices, eliminating the need to transmit raw data.  We validated \textit{FedJam} through extensive simulations and real-world testbed deployment, showing up to 15\% higher detection accuracy than state-of-the-art baselines, while requiring 60\% fewer communication rounds to converge, with low overhead across computation, memory, and communication.

\clearpage
\bibliographystyle{IEEEtran}
\bibliography{references}

\end{document}

%% file: table_iid.tex
\begin{table}[t]
\centering
\small
\caption{\textbf{Detection accuracy (\%) at convergence. }}
\label{tab:accuracy_pivot}
\setlength{\tabcolsep}{6pt}
\renewcommand{\arraystretch}{0.9}
\begin{tabular}{lccc}
\toprule
\textbf{Method} & \textbf{10 Clients} & \textbf{25 Clients} & \textbf{50 Clients} \\
\midrule
FewShotDense~\cite{federated_learning_spectrogram_5}    & 95.10 {\scriptsize$\pm$ 0.31} & 88.10 {\scriptsize$\pm$ 0.24} & 84.10 {\scriptsize$\pm$ 0.39} \\
SpectroNet~\cite{spectrogram_1}      & 96.00 {\scriptsize$\pm$ 0.27} & 93.80 {\scriptsize$\pm$ 0.34} & 89.00 {\scriptsize$\pm$ 0.25} \\
JamShield~\cite{jamshield}       & 58.40 {\scriptsize$\pm$ 0.40} & 50.70 {\scriptsize$\pm$ 0.21} & 46.70 {\scriptsize$\pm$ 0.28} \\
1D-CNN~\cite{kiranyaz20211d}            & 46.60 {\scriptsize$\pm$ 0.33} & 42.53 {\scriptsize$\pm$ 0.30} & 40.89 {\scriptsize$\pm$ 0.29} \\
TinyViT~\cite{wu2022tinyvit}         & 89.60 {\scriptsize$\pm$ 0.87} & 84.60 {\scriptsize$\pm$ 0.52} & 77.10 {\scriptsize$\pm$ 0.92} \\
SpectroKPI-Fuser & 56.10 {\scriptsize$\pm$ 0.25} & 45.40 {\scriptsize$\pm$ 0.37} & 40.50 {\scriptsize$\pm$ 0.32} \\
\textbf{FedJam} & \textbf{99.20 {\scriptsize$\pm$ 0.22}} & \textbf{96.70 {\scriptsize$\pm$ 0.30}} & \textbf{92.60 {\scriptsize$\pm$ 0.27}} \\

\bottomrule
\end{tabular}
\end{table}

%% file: table_non_iid.tex
\begin{table}[t]
\centering
\small
\caption{\textbf{Detection accuracy (\%) at convergence for Non-IID settings}. Columns “2” and “3” indicate the number of classes included in the training data of each client, selected from a total of 4 classes.}
\label{tab:non_iid_grouped}
\setlength{\tabcolsep}{3pt}
\renewcommand{\arraystretch}{0.9}
\begin{tabular}{lcccccc}
\toprule
\textbf{ } & \multicolumn{2}{c}{\textbf{10 Clients}} & \multicolumn{2}{c}{\textbf{25 Clients}} & \multicolumn{2}{c}{\textbf{50 Clients}} \\
\cmidrule(lr){2-3} \cmidrule(lr){4-5} \cmidrule(lr){6-7}
\multirow{-2}{*}{\vspace{-7pt}\textbf{Non-IID}}  & \textbf{2} & \textbf{3} & \textbf{2} & \textbf{3} & \textbf{2} & \textbf{3} \\
\midrule
FewShotDense & 81.12 & 92.76 & 71.99 & 87.69 & 72.78 & 81.14 \\
SpectroNet   & 84.15 & 95.00 & 81.08 & 92.21 & 75.68 & 85.12 \\
JamShield    & 46.29 & 51.67 & 44.26 & 47.46 & 38.29 & 45.58 \\
SpectroKPI-Fuser    & 48.69 & 54.67 & 46.36 & 48.96 & 36.79 & 41.21 \\
\textbf{FedJam}       & \textbf{96.38} & \textbf{98.97} & \textbf{91.28} & \textbf{96.18} & \textbf{86.58} & \textbf{90.57}  \\
\bottomrule
\end{tabular}
\end{table}